\documentclass[fleqn,usenatbib]{mnras}

\usepackage[T1]{fontenc}

\DeclareRobustCommand{\VAN}[3]{#2}
\let\VANthebibliography\thebibliography
\def\thebibliography{\DeclareRobustCommand{\VAN}[3]{##3}\VANthebibliography}

\usepackage{graphicx}	
\usepackage{amsmath}	
\usepackage{amssymb}	

\title[From the CMF to the IMF]{From the CMF to the IMF: Beyond the Core-Collapse Model}

\author[V.-M. Pelkonen et al.]{
V.-M. Pelkonen,$^{1}$\thanks{E-mail: veli.matti.pelkonen@icc.ub.edu}
P. Padoan,$^{1,2}$
T. Haugb{\o}lle$^{3}$
and \AA. Nordlund$^{3}$
\\
$^{1}$Institut de Ci\`{e}ncies del Cosmos, Universitat de Barcelona, IEEC-UB, Mart\'{i} i Franqu\'{e}s 1, E08028 Barcelona, Spain\\
$^{2}$ICREA, Pg. Llu\'{i}s Companys 23, 08010 Barcelona, Spain\\
$^{3}$Niels Bohr Institute, University of Copenhagen, {\O}ster Voldgade 5-7, DK-1350 Copenhagen, Denmark
}

\date{Accepted XXX. Received YYY; in original form ZZZ}

\pubyear{2020}

\begin{document}
\label{firstpage}
\pagerange{\pageref{firstpage}--\pageref{lastpage}}
\maketitle

\begin{abstract}
Observations have indicated that the prestellar core mass function (CMF) is similar to the stellar initial mass function (IMF), except for an offset towards larger masses. This has led to the idea that there is a one-to-one relation between cores and stars, such that the whole stellar mass reservoir is contained in a gravitationally-bound prestellar core, as postulated by the core-collapse model, and assumed in recent theoretical models of the stellar IMF. We test the validity of this assumption by comparing the final mass of stars with the mass of their progenitor cores in a high-resolution star-formation simulation that generates a realistic IMF under physical conditions characteristic of observed molecular clouds. Using a definition of bound cores similar to previous works we obtain a CMF that converges with increasing numerical resolution. We find that the CMF and the IMF are closely related in a statistical sense only; for any individual star there is only a weak correlation between the progenitor core mass and the final stellar mass.  In particular, for high mass stars only a small fraction of the final stellar mass comes from the progenitor core, and even for low mass stars the fraction is highly variable, with a median fraction of only about 50\%. We conclude that the core-collapse scenario and related models for the origin of the IMF are incomplete. We also show that competitive accretion is not a viable alternative.

\end{abstract}



\section{Introduction}

The origin of the stellar initial mass function (IMF) is still considered an open problem, partly due to a lack of observational constraints. While the overall shape of the IMF is well documented in stellar clusters, its possible dependence on environmental parameters is still unclear. The turbulent nature of star-forming gas compounds the problem, although it may also be viewed as a key to solve it \citep[e.g.][]{Padoan+Nordlund02imf,Hennebelle+Chabrier08imf,Hopkins12imf}. 

The similarities of the stellar IMF \citep{Kroupa01,Chabrier05} with the prestellar core mass function (CMF) derived from simulations \citep[e.g.][]{Klessen2001,Tilley+Pudritz04,Tilley+Pudritz07,Padoan+07imf,Chabrier+Hennebelle2010,Schmidt+2010} and observations \citep[e.g.][]{Motte+98,Alves+07,Enoch+07,Nutter+Ward-Thompson07,Konyves+10,Konyves+15,Marsh+16,Sokol+19,Konyves+2020_orion,Ladjelate+2020_ophiucus} has led to the suggestion that prestellar cores are true progenitors of stars, meaning that the mass reservoir of a star is fully contained in the progenitor core. The final mass of a star, $M_{\rm star}$, is then given by that of its progenitor core, $M_{\rm prog}$, multiplied by an efficiency factor, $M_{\rm star}=\epsilon_{\rm prog} M_{\rm prog}$, and the IMF is interpreted as a CMF offset in mass by the factor $\epsilon_{\rm prog}$ \citep[e.g.][]{Alves+07,Andre+10,Andre+14}.

The idea that in the prestellar phase the stellar mass reservoir is fully contained in a bound overdensity is a fundamental assumption in the IMF models of \citet{Hennebelle+Chabrier08imf} and \citet{Hopkins12imf}. A bound overdensity collapses when it becomes gravitationally unstable, and some form of thermal, magnetic or kinetic support is needed before the start of the collapse, if the overdensity is to accumulate all the stellar mass reservoir. Given the very broad mass range of stars, and the relatively small temperature variations in the dense molecular gas, magnetic and/or kinetic energy support are needed, particularly in the case of the most massive stars that would require the most massive prestellar cores \citep{McKee+Tan02,McKee+Tan03}. This scenario for the origin of the IMF and massive stars, often referred to as \emph{core collapse}, has been called into question by results of numerical simulations and observations.

Simulations of star formation in turbulent clouds result in relatively long star-formation timescales: high-mass protostars acquire their mass on a timescale comparable to the dynamical timescale of the star-forming cloud \citep{Bate09,Bate12,Padoan+14acc,Padoan+20SN_massive}, which is much longer than the characteristic free-fall time of prestellar cores. In the case of the formation of intermediate- and high-mass stars, using SPH simulations or tracer particles in grid simulations, it has been shown that the mass reservoir of sink particles extends far beyond the prestellar cores \citep{Bonnell+2004,Smith+09massive,Padoan+20SN_massive}. Both results seem to be at odds with the core-collapse idea. 

Recent interferometric observations of regions of massive star formation have revealed a scarcity of massive prestellar cores that could serve as the mass reservoir for high-mass stars 
\citep[e.g.][]{Sanhueza+17,Sanhueza+19,Li+19,Pillai+19,Kong19cmf,Servajean+19}, implying that such a mass reservoir is spread over larger scales. This is also suggested by the evidence of parsec-scale filamentary accretion onto massive protostellar cores \citep[e.g.][]{Peretto+13}, as well as in smaller scale observations of a flow originating from outside the dense core of a protostar \citep{Pineda+20}, which has also been seen in simulations \citep{Kuffmeier+17} and studied in semi-analytical models \citep{Dib+2010}. \citet{Chen+19} report observations of non-gravitationally bound cores potentially formed by pressure compression and turbulent processes, which by definition involves accretion from outside of a gravitationally bound core. Furthermore, while the prestellar CMFs in the Aquila and Orion regions are found to peak at a mass a few times larger than that of the stellar IMF, corresponding to a progenitor core efficiency $\epsilon_{\rm prog}\sim 0.2-0.4$ \citep{Andre+10,Konyves+15,Konyves+2020_orion}, the peak mass of the CMFs in Taurus and Ophiucus is very close to that of the stellar IMF, or even smaller if candidate cores are included \citep{Marsh+16,Ladjelate+2020_ophiucus}. This would imply that $\epsilon_{\rm prog}\ge 1$ in Taurus and Ophiucus, contrary to the core-collapse scenario. Protostellar jets and outflows remove a significant fraction of the accreting mass \citep[e.g.][]{Matzner+McKee2000,Tanaka+17}, so a core-collapse model requires $\epsilon_{\rm prog}\lesssim 0.5$. 

Some of the alternatives to the core-collapse models are the competitive accretion model \citep[e.g][]{Zinnecker82,Bonnell+2001a,Bonnell+2001b,Bonnell+Bate2006} and the inertial-inflow model \citep{Padoan+20SN_massive}. In the competitive accretion model, all stars have initially low mass, and most of their final mass is accreted after the initial collapse and may originate far away from the initial core. The accretion is due to the stellar gravity, so the accretion rate is expected to grow with the stellar mass. However, the Bondi-Hoyle accretion rate in turbulent clouds is too low for this model to explain the stellar IMF, unless the star-forming region has a very low virial parameter, with a high density and comparatively low velocity dispersion \citep{Krumholz+05nature}. Thus, although this model is quite different from the core-collapse model, it still requires that the feeding region of a star is gravitationally bound and essentially in free-fall collapse. 

In the inertial-inflow model \citep{Padoan+20SN_massive}, stars are fed by mass inflows that are not driven primarily by gravity, as they are an intrinsic feature of supersonic turbulence. The scenario is inspired by the IMF model of \citet{Padoan+Nordlund02imf}, where prestellar cores are formed by shocks in converging flows. The characteristic time of the compression is the turnover time of the turbulence on a given scale, which is generally larger than the free-fall time in the postshock gas, so a prestellar core may collapse into a protostar well before the full stellar mass reservoir has reached the core \citep{Padoan+Nordlund11imf}\footnote{The
distinction between the prestellar CMF and the stellar IMF based on the difference between the turbulence turn-over time and the free-fall time was not mentioned in \citet{Padoan+Nordlund02imf}, and the model has been presented and interpreted incorrectly as a core-collapse model. The implication of the model with respect to the distinction between the prestellar CMF and the stellar IMF was explained later in \citep{Padoan+Nordlund11imf}.  
}. 
After the initial collapse, the star can continue to grow, as the converging flows that were feeding the prestellar core continue to feed the star, through the mediation of a disk. Thus, the stellar mass reservoir can extend over a turbulent and \emph{unbound} region much larger than the prestellar core. Because converging flows occur spontaneously in supersonic turbulence and can assemble the stellar mass without relying on a global collapse or on the gravity of the growing star \citep{Padoan+14acc,Padoan+20SN_massive}, the inertial-inflow scenario is fundamentally different from competitive-accretion.

The main goal of this work is to test the core-collapse model, hence the validity of the IMF and massive-star formation theories based on that scenario. 
Even though there are many possible observational or theoretical definitions of a prestellar core, our focus is a definition that stems directly from the core-collapse model, where a fundamental assumption is that a prestellar core is a gravitationally unstable object containing the whole mass reservoir of the star.
The most direct way to test the core-collapse model with numerical experiments is to use simulations where the formation and growth of stars is captured with accreting sink particles, hereafter called 'stars' or 'star particles'. The final mass of each star particle, $M_{\rm star}$, can be compared with the mass of its progenitor core, $M_{\rm prog}$, and the hypothesis of the core-collapse model, $M_{\rm star}\approx\epsilon_{\rm prog} M_{\rm prog}$ (with $\epsilon_{\rm prog}\lesssim 0.5$ and approximately independent of mass), can be tested directly.

The relation between core and star particle masses in the Salpeter range of masses was investigated by \citet{Smith+09} and by \citet{Gong+Ostriker15}. Both works found a reasonable correlation between the masses of cores and the masses of sink particles after one local free-fall time of each core, when the sink particles were still growing. At later times, the sink particle masses were significantly larger than the core masses (see Figure~10 in \citet{Smith+09} and Figure~22 in \citet{Gong+Ostriker15}), although \citet{Chabrier+Hennebelle2010} suggested that feedback, neglected in those works, could halt the later accretion and thus preserve the statistical correlation. These early results, where the sink particle masses develop to significantly exceed the core masses, are in conflict with the core-collapse model, as they imply that $\epsilon_{\rm prog} > 1$.  However, both simulations had relatively low resolution, and therefore could not resolve the peak of the IMF \citep{Haugbolle+18imf}. Furthermore, because a large fraction of sink particles were still accreting at the end of the simulations, it is unclear what the final correlation between sink particle and core masses should be. The absence of driving (from larger scales or from artificial forcing) results in rapidly decaying turbulence, which together with the small dynamic range of the simulations means that velocity fluctuations at core scales must become severely underestimated at late times. In addition, both works neglected magnetic fields, which are known to play an important role in the fragmentation process \citep[e.g.][]{Padoan+07imf,Krumholz+Federrath19,Ntormousi+Hennebelle19}.

Higher-resolution simulations were used by \citet{Mairs+14} to study the CMF without magnetic fields, and by \citet{Smullen+20} including magnetic fields. Cores were selected from two-dimensional maps with CLFIND2D \citep{Williams+94} to mimic the observations in \citet{Mairs+14}, and from the three-dimensional density field as dendrogram leaves in \citet{Smullen+20}. Both works found that the final sink particle masses were often significantly larger than the initial masses of the corresponding prestellar cores, in agreement with \citet{Smith+09} and \citet{Gong+Ostriker15} and contrary to the core-collapse model. Despite the larger resolution, these works did not produce realistic stellar IMFs, in part because the resolution was still insufficient, and in part because the gas accretion onto the sink particles was assumed to have an efficiency of 100\%. Furthermore, the CMF in \citet{Smullen+20} was computed at a much lower resolution than the actual simulation. Neither the peak of the IMF nor that of the CMF were shown to be numerically converged, so the two functions could not be accurately compared. Finally, none of the four works discussed above used tracer particles, so the origin of the stellar mass reservoir could not be tracked backward in time \citep[though it could have been done with the SPH simulation in][]{Smith+09}.

To more realistically investigate the relation between the masses of stars and their progenitor cores we use one of the adaptive-mesh-refinement (AMR), magneto-hydrodynamic (MHD) simulations of \citet{Haugbolle+18imf}. While other simulations have reproduced the stellar IMF including different combinations of physical processes \citep[see review by][]{Lee+20}, this was the first work to demonstrate that isothermal, supersonic, MHD turbulence alone can produce a numerically converged IMF. The high dynamic range, the inclusion of magnetic fields, a sub-grid model of the effects of outflows, and the long integration time overcome the limitations of previous studies, and result in a mass distribution of star particles that is consistent with the observed stellar IMF, from brown dwarfs to massive stars. By the end of this simulation, a large fraction of star particles has stopped accreting, so final masses are well defined. Stars stop accreting when their converging mass inflows are terminated, when pressure forces decouple the gas from the stars, or when they are dynamically ejected from multiple systems while still accreting.

The paper is structured as follows. \S~2 briefly summarizes the simulation and the numerical methods, and \S~3 describes the selection of bound progenitor cores around newly formed star particles from the simulation. \S~4 describes the physical properties of the progenitor cores, including their masses, sizes and sonic Mach numbers, and verifies that the progenitors are supercritical with respect to their magnetic support. The final star particle masses are compared with their progenitor core masses in \S~5, and tracer particles are used to identify the fractions of mass accreted from the progenitor and from farther away. The results are discussed in \S~6 in the context of the core collapse model and the observations of the core mass function. \S~7 summarizes the conclusions.

\section{Simulation}

The simulation used in this study is the \emph{high} reference simulation from \cite{Haugbolle+18imf} (see their Table~1). Details of the numerical methods and the simulation setup are in that paper, and are only briefly summarized in the following. 

The MHD code used for the simulation is a version, developed in Copenhagen, of the public AMR code RAMSES \citep{Teyssier07}. It includes random turbulence driving and a robust algorithm for star particles. Periodic boundaries and an isothermal equation of state are adopted for the simulation. The grid refinement is based on overdensity, first at ten times the mean density and then at each factor of four in density. With a root grid of $256^3$ cells and six levels of refinement, the highest effective resolution is 16,384$^3$, corresponding to a smallest cell size of 50~AU for the assumed box size of 4~pc. The initial turbulent state is achieved by running the simulation without self-gravity for $\sim 20$ dynamical times, with a random solenoidal acceleration giving an rms sonic Mach number of approximately 10. Then self-gravity is switched on while maintaining the random driving, and the simulation is run for another $\sim 2.6$~Myr, with snapshots saved every 22~kyr.

The physical parameters of the simulation are as follows. The temperature is set at 10~K and the mean molecular weight $\mu= 2.37$, appropriate for cold molecular clouds, resulting in an isothermal sound speed of $0.18 \; \rm km s^{-1}$. Assuming a box size $L_{\rm box} = 4$~pc, the total mass, mean density and mean magnetic field strength are $M_{\rm box}=3000 \,M_\odot$, $\bar{n}=795 \, \rm cm^{-3}$, and  $B_{\rm box}=7.2 \, \mu \rm G$. The corresponding free-fall time and dynamical time are 1.18~Myr and 1.08~Myr, respectively.

At the end of the simulation, 413 star particles have been created, with a mass distribution consistent with the observed stellar IMF \citep{Haugbolle+18imf}. As discussed in \citet{Haugbolle+18imf} and further demonstrated in Appendix~\ref{app_imf}, the peak of the IMF converges to a well-defined value. Star particles are created at a local gravitational potential minimum when the gas density in the cell is above a critical density, set at $1.7\times 10^9 \, \rm cm^{-3}$. They are created without mass, but start accreting from their surroundings. 

The mass-loss from winds and outflows and their feedback on the surrounding gas is not modelled in detail in the current simulation. However, the protostellar feedback processes are accounted for through an accretion efficiency factor, $\epsilon_{\rm acc} = 0.5$, meaning that only half of the accreting mass is added to the stellar mass. This is consistent with the reduction in stellar masses found in the small-scale simulations by \citet{Federrath+14}, where outflows are included. Their outflow subgrid model assigns 30\% of the accreting mass to the outflows. In addition, the outflows disperse part of the surrounding gas, which further reduces the global star formation rate by another 20\%.\footnote{The global reduction in the stellar masses in \citet{Federrath+14} is actually a factor of three, due to the increase in the number of stars when outflows are included. A similar reduction factor is obtained also with $\epsilon_{\rm acc} = 0.5$ in our simulations (compared with equivalent runs where $\epsilon_{\rm acc} = 1.0$), due to the reduced gravity of the sink particles.} Thus, the inclusion of $\epsilon_{\rm acc}$ in our simulation functions as a sub-grid model for protostellar feedback that effectively includes both the direct mass-loss from the outflows and the reduction in the accretion rate due to the feedback on the surrounding gas.

Despite the 50\% of the accreting mass being removed rather than launched back into the simulation in our subgrid model, the star formation timescales are on average of the order of the observational estimates. As shown in Figure 11 of \citet{Haugbolle+18imf}, the stars in our simulation accrete 95\% of their final mass in a time of $0.51(M_{\rm star}/M_{\odot})^{0.58}$~Myr (the mean and median time, for all stars that have finished accreting by the end of the simulation, are 0.38 and 0.22~Myr, respectively), comparable to the observational value of $\sim 0.5$~Myr for the sum of the Class 0 and I phases \citep[e.g.,][]{Enoch+09,Evans+09,Dunham+14}. This agreement suggests that the main role of outflows is to reduce the accretion efficiency (as in our subgrid model) rather than shortening the star-formation time by halting the accretion completely. 

We use $10^8$ tracer particles in the global simulation to trace the mass flow in the box. The tracer particles are passively advected with the fluid velocity. This may lead to a growing inaccuracy between the gas density field and the density field defined by the tracer particles \citep{Genel+13}.  This is a secular effect that at the end of the simulation, after 2.6~Myr, may introduce a discrepancy between mass defined by tracers compared to defined by the gas mass of $\approx$10\% on the scale of 1 $M_\odot$. We remedy this effect by redefining trace particle masses according to the density field, at the time of core selection, making the core masses defined by gas density or by tracers consistent. The typical mass accretion time scale for a low-mass core is 100 kyr, and the difference between using the gas density field and passive tracers is below the percent-level in the analysis below. Passive tracers are accreted probabilistically to star particles with a probability matching the fraction of gas accreted in the enclosing cells.

\begin{figure}
   \centering
   \includegraphics[width=\hsize]{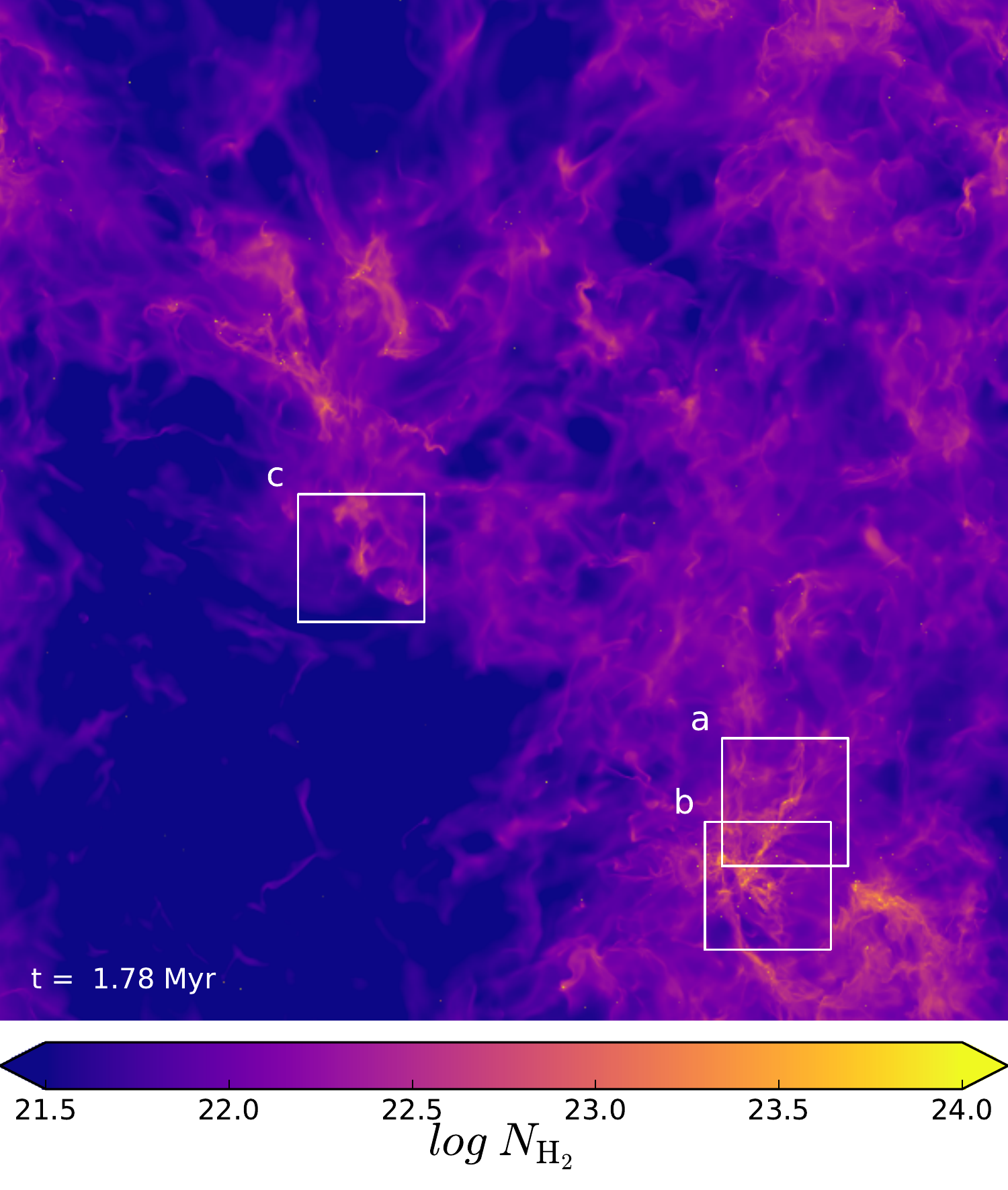}
   \includegraphics[width=\hsize]{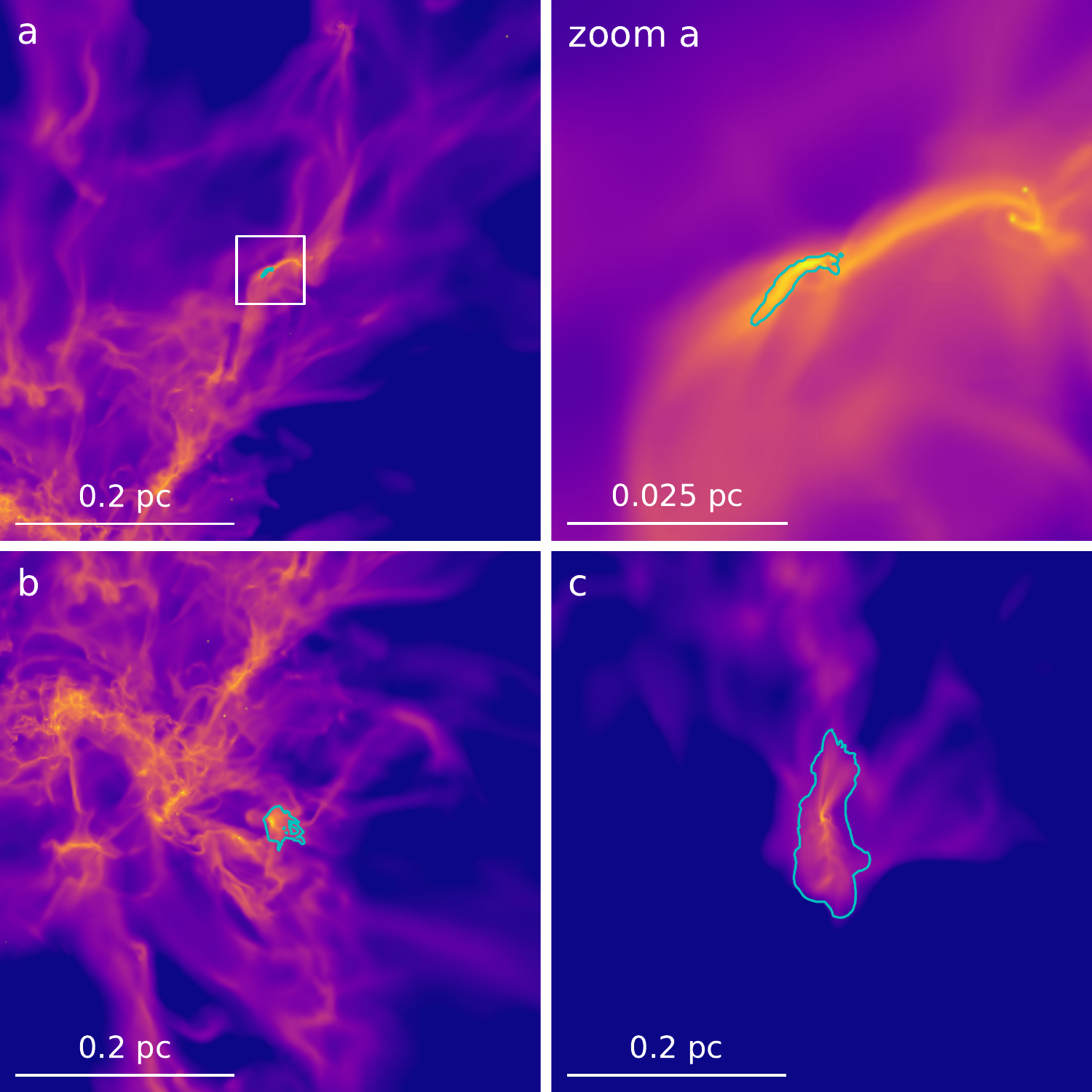}   
      \caption{{\it Upper panel}: Column density of a snapshot of the whole 4~pc simulation in logarithmic scale. The three white boxes mark the position of the zoom-in regions shown in the lower panel. Stars are visible as one-pixel density enhancements (see \S~\ref{detection}). {\it Lower panel }: Three 0.5~pc zoom-in regions, centred on a newly-born star, in the same color scale as in the upper panel. Cyan contours are drawn where there is at least one cell belonging to the progenitor core on the line of sight. In the case of core a, an additional zoom-in is shown in the upper-right frame, due to the small core size.}
         \label{columndensity}
   \end{figure}

Figure~\ref{columndensity} shows a snapshot of column density of the whole simulation box. The three zoom-in regions show examples of 0.5~pc subcubes that are used in the detection of the progenitor cores (see Section~\ref{detection}), drawn in the same color scale. 

\section{Selection of Progenitor Cores}

\subsection{Clumpfind algorithm} 

Cores are selected from the simulation with the clumpfind algorithm introduced in \citet{Padoan+07imf}, updated in this work to consider both thermal and kinetic energies. In the clumpfind algorithm, cores are defined as ``gravitationally unstable connected overdensities that cannot be split into two or more overdensities of amplitude $\delta n / n > f$, all of which remain unstable''. Cores are considered unstable if their gravitational binding energy is larger than the sum of their kinetic and thermal energies (see \S~3.2 for details). The algorithm scans the density field with discrete density levels, each of amplitude $f$ relative to the previous one. Only the connected regions above each density level that are found to be unstable are retained. After this selection, the unstable cores from all levels form a hierarchy tree. Only the final (unsplit) core of each branch is retained. Each core is assigned only the mass within the density isosurface that defines the core (below that density level the core would be merged with its next neighbor). This algorithm is different from the construction of a dendrogram \citep{Rosolowsky+08}, where a hierarchy tree of cores is first computed irrespective of the cores energy ratios, and unstable cores are later identified among the leaves of the dendrogram. In our algorithm, the condition for instability is imposed while building the tree, otherwise some large unstable cores would be incorrectly eliminated if they were split into smaller cores that were later rejected by being stable.

Once the physical size and mean density of the system are chosen, the clumpfind algorithm depends only on three parameters: (1) the spacing of the discrete density levels, $f$, (2) the minimum density above which cores are selected, $n_{\rm min}$, and (3) the minimum number of cells per core. In principle, there is no need to define a minimum density, but in practice it speeds up the algorithm. The parameter $f$ may be chosen according to a physical model providing the value of the smallest density fluctuation that could collapse separately from its contracting background. In practice, the $f$ parameter is set by looking for numerical convergence of the mass distribution with decreasing values of $f$. The minimum number of cells is set to 4, to ensure that all the detected cores are at least minimally resolved in the simulation.

\subsection{Stability condition}
\label{stability}

In the clumpfind algorithm, the absolute value of the gravitational binding energy is given by the formula for a uniform sphere:
\begin{equation}
E_{\rm g} = \frac{3 G M_{\rm prog}^2}{5 R_{\rm prog}},
\label{eqeg}
\end{equation}
where $G$ is the gravitational constant, $M_{\rm prog}$ is the mass of the core, $R_{\rm prog}$ is the radius of a sphere of an equivalent volume to that of the core. This proved to be a good approximation for the gravitational binding energy for the cores that we find, when compared to a pair-wise calculation of the potential energy between the cells of the core. 

The kinetic energy is given by:
\begin{equation}
E_{\rm k} = \sum_{\rm i=1}^N \frac{1}{2}m_{\rm i} (v_{\rm i} - \bar{v})^2,
\end{equation}
where N is the number of cells in the core, $m_{\rm i}$ and $v_{\rm i}$ are the mass and the velocity of each cell, and $\bar{v}$ is the mass-weighted mean velocity of the core. The thermal energy is given by the formula for an ideal di-atomic gas:
\begin{equation}
E_{\rm t} = \frac{5}{2} N k T,
\end{equation}
where $N$ is the number of hydrogen molecules, $k$ is the Boltzmann constant and $T$ is the temperature, set to 10K.
The condition that cores are unstable is then based on the energy ratio,
\begin{equation}
\frac{E_{\rm k} + E_{\rm t}}{E_{\rm g}} \leq 1.
\label{eqalphavire}
\end{equation}

The condition should be $(E_{\rm k} + E_{\rm t})/E_{\rm g} \leq 1/2$, based on the virial theorem, if surface terms were included. Our choice implicitly assumes a Bonnor-Ebert model including surface terms of the order of half the internal energy. To test this approximation, we have calculated the surface terms (see Appendix~\ref{app_surface}). The median ratio of the surface terms divided by the internal energy is about 0.5. Thus, our $\alpha_{\rm vir} = 1$ without surface terms is on average equal to $\alpha_{\rm vir} = 0.5$ with surface terms accounted for, as in the virial theorem. \citet{Smith+09} and \citet{Gong+Ostriker15} use the same condition as in Eq.~\ref{eqalphavire} to select unstable cores in their simulations.

\subsection{Selection setup}
\label{detection}

We extract $1,024^3$ subcubes of 0.5~pc in size centred on each star particle, in the snapshot where the star particle first appears. This resolution corresponds to $8,192^3$ resolution for the full 4~pc simulation box and a physical scale of 100~AU per cell. As found previously by e.g.~\citet{Hennebelle2018}, the numerical resolution can have a large effect on the masses of the selected cores; lower resolutions can result in higher core masses as well as increase the number of cores with multiple stars inside them. 

Nevertheless, given our large underlying resolution, on top of which we only vary the resolution of the clumpfind algorithm, the peak of the CMF changes much less than in e.g.~\citet{Hennebelle2018}, and the peak converges towards a definite value as the resolution increases (see Appendix~\ref{app_parameters}). The parameters of the clumpfind algorithm and the effect of numerical resolution are discussed in Appendix~\ref{app_parameters}.

The clumpfind algorithm should in principle be applied to the whole periodic box, as the cores could depend on how neighboring cores are selected, but this would be computationally infeasible at the highest resolution. However, most of the cores are well-defined within a 0.5~pc subcube. Due to their large size, for 22 cores with masses higher than $5 \; M_\odot$ in the 0.5~pc subcube, we repeat the clumpfind search using 1~pc subcubes.

Ideally, we would use a snapshot at a time immediately before the central star is created. In the absence of such a snapshot, and in the spirit of characterizing the core condition at the time immediately before the star formation, we add twice the initial mass of the central star particle ($M_{\rm sink}/\epsilon_{\rm acc}$) back to the density cube to account for the gravitational energy of the gas that has already been accreted onto or ejected by the central star particle between the time of its formation and the time of the detection snapshot. The mass of other stars present in the subcube is also added to the density cube, using a factor of 2 if they formed in the same snapshot as the central star or a factor of 1 if they had formed in a previous snapshot (we assume that the jet would have time to disperse the $\epsilon_{\rm acc}$ fraction of the accreted mass away from the core in that case). The mass is added into the cell where each star is located. The details of how the mass is added do not affect the calculation of the gravitational binding energy, as only the total mass is needed in the uniform sphere approximation of Equation~\ref{eqeg}. Additionally, this method ensures that, if there is no extended material around the star, the single cell 'core' is not picked up, because the minimum number of cells per core is set to 4. 

For the overdensity amplitude, $f$, we adopt a conservative value, $f = 2\%$. Given that the 0.5~pc subcubes are in overdense regions of the full box, we select the minimum number density level to be $n_{\rm min} = 10\,\bar{n} = 7950\,\textrm{cm}^{-3}$. We have varied $f$ and $n_{\rm min}$ using $f = 8\%$ and $n_{\rm min} = 2\,\bar{n}$ and verified it resulted in almost identical core masses.

The clumpfind algorithm is run for each subcube, to find all the unstable cores. Then a search is made for the core that contains the star particle (in the central cell of the subcube). This core is labelled as the progenitor core for that star particle. If no core containing the central cell is found, we search for the closest core inside a $21^3$ cube around the star, and record the distance to the star if a core is found. Otherwise, the distance is set to a high number to indicate that no match is found.

\section{Progenitor Cores}
\label{sectprog}

The clumpfind algorithm outputs a list of cell indices that are part of each core, as well as the masses of the cores. Once the progenitor cores are identified, their cell indices are used to find any other star particles within the progenitor. As explained in Section~\ref{detection}, we include twice the masses of the star particles that formed in the same snapshot as the primary star into the density cube, whereas older star particles were already considered to be stars.

Out of a total of 413 star particles in the simulation, we are able to match 382 unambiguously with a progenitor core. Among the other 31 stars, predominantly of low mass, 22 of them are found within 1,000~AU of a core and 9 are more distant. Some of these stars are brown dwarfs that have collapsed quickly from a small core and have already accreted their parent core by the time of the snapshot we analyze. Others are stars that have already decoupled from the accreting gas due to pressure forces acting on the gas or dynamical encounters with other stars. A small fraction of them may be numerical artifacts, as discussed in \citet{Haugbolle+18imf}. We exclude those 31 stars from the analysis.

We further divide our 382 unambiguously matched progenitors in three categories based on the stars within the progenitor: 1) a single star (the primary one, 312 cores); 2) multiple stars, but all formed in the same snapshot (32 cores); 3) one or more older stars in addition to the primary one (38 cores). The older stars may dominate the mass reservoir inside the core. To be conservative, we exclude category three from the analysis.

The final sample used in the paper is therefore 344 stars with an unambiguous progenitor detection and no older stars inside the progenitor core. 

To study the mass distribution of the accreted gas (see Section~\ref{progvssink}), we extract the passive tracer particles that are within a progenitor core at the time of its identification. These tracer particles are used to calculate the fractions of the progenitor mass accreted by the primary star ($f_{\rm tr,prog}$), accreted by any other stars ($f_{\rm tr,other}$), and not accreted by any star ($f_{\rm tr,unacc}$) at the end of the simulation. $f_{\rm tr,prog}$ is defined as $M_{\rm tr,prog}/M_{\rm prog}$, where $M_{\rm tr,prog}$ is the mass of all the tracers inside the progenitor that will be accreted by the primary star. The mass fraction going to other stars is simply defined as tracer particles that will be accreted by any other stars by the end of the simulation; these other stars do not need to form at the same time or become binaries, and most often do not. Progenitors that do form multiple stars in the same snapshot are flagged as such, as explained in the previous paragraph. Furthermore, we calculate the fraction of the final stellar mass contained within the progenitor core: $f_{\rm tr, star} = \epsilon_{\rm acc} f_{\rm tr,prog} M_{\rm prog} / M_{\rm star}$. Here, the efficiency factor $\epsilon_{\rm acc}$ is already included in the final stellar mass, $M_{\rm star}$, and thus needs to be applied for the tracer mass as well. These fractions evolve while the stars accrete their mass, and some of them have not stopped accreting yet. Thus, we further check if the stars have stopped accreting at the end of the simulation by computing the final accretion rate as the mass increase between the last two snapshots, divided by the time interval between the snapshots. This is an accretion rate averaged over 22~kyr. If the stars need 10~Myr or longer to double their final mass given that accretion rate, we consider them as having finished accreting (242 stars out of 344). Finally, to characterize the size of the region from which a star accretes its mass, we calculate the inflow radius, $R_{\rm 95}$, as was done in \citet{Padoan+20SN_massive}. $R_{\rm 95}$ is defined as the radius of the sphere that contains 95\% of the tracer particles' mass accreted by the star by the end of the simulation. We refer to it as the inflow radius in reference to our scenario where growing stars are fed by inertial inflows, as extensively demonstrated in \citet{Padoan+20SN_massive}. The mass fractions and the inflow radius defined through the tracer particles are used in \S~\ref{progvssink}, where we address the relationship between the stars and their progenitor cores. 

   \begin{figure}
   \centering
   \includegraphics[width=\hsize]{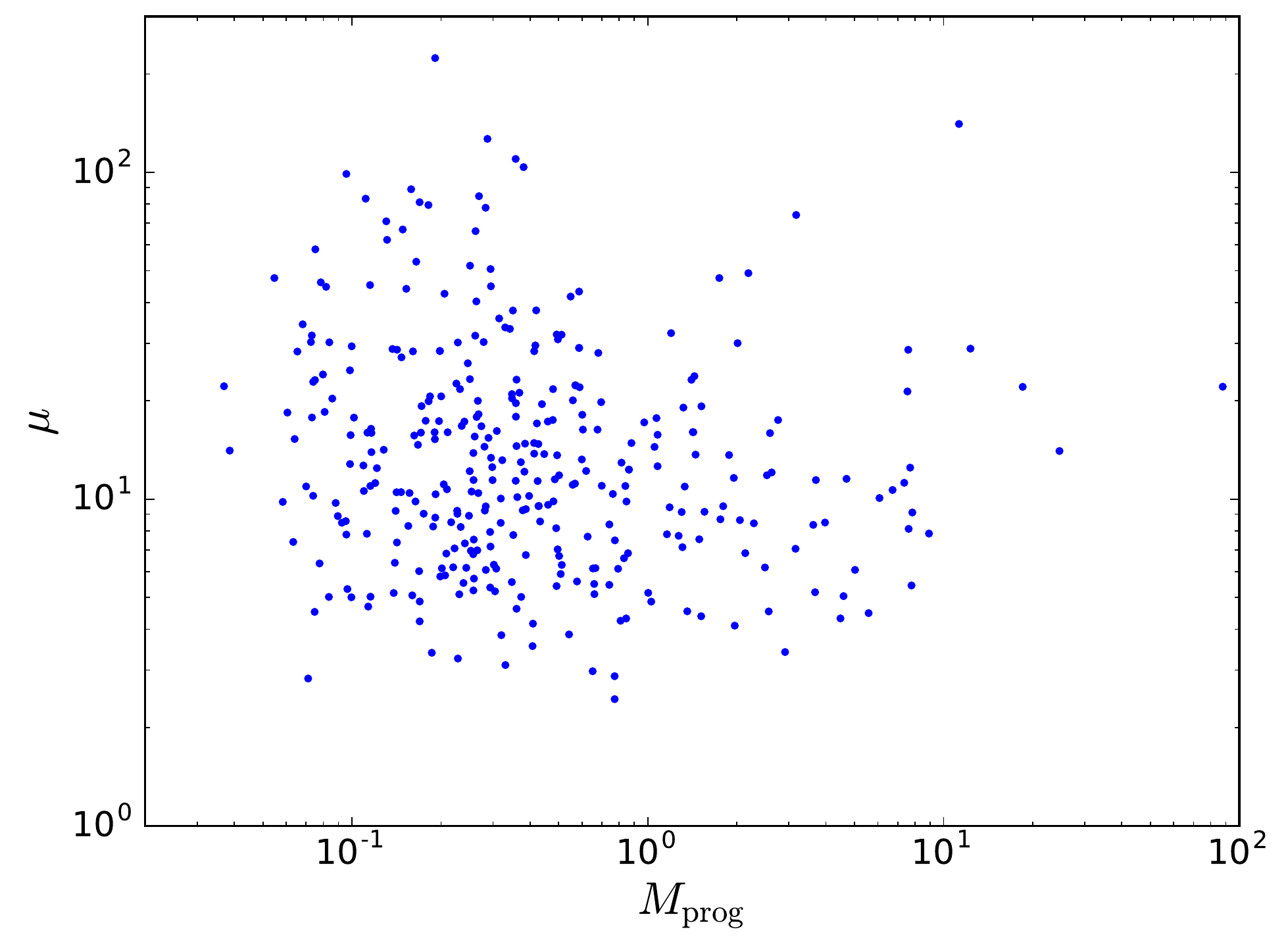}
      \caption{Normalized mass-to-flux ratio, $\mu$, as a function of the progenitor mass, $M_{\rm prog}$. All progenitor cores are magnetically supercritical ($\mu>1$), as expected for collapsing cores in the simulation.
              }
         \label{masstoflux}
   \end{figure}

Since the clumpfind algorithm does not consider the magnetic field, we check for magnetic support against gravitational collapse after the fact. We follow the same methodology as in \citet{Ntormousi+Hennebelle19}. We calculate the magnetic critical mass \citep{Mouschovias+Spitzer76},
\begin{equation}
M_{\rm c\phi} = \frac{c_1}{3\pi} \left(\frac{5}{G}\right)^\frac{1}{2} \phi
\end{equation}
where $c_1 = 1$ for a uniform sphere, $G$ is the gravitational constant and $\phi$ is the magnetic flux. 

We measure $\phi$ for each core by selecting three planes going through the core centre perpendicular to each axis. We then flag the cells belonging to the core, and sum the flux along the perpendicular axis over all the cells. We then select the highest of the three fluxes to calculate the magnetic critical mass. The mass-to-flux ratio, normalized by the magnetic critical mass, is then simply given by:
\begin{equation}
\mu = M_{\rm prog} / M_{\rm c\phi},
\end{equation}
where $M_{\rm prog}$ is the mass of the core and $M_{\rm c\phi}$ is the magnetic critical mass, assuming a uniform sphere. Figure~\ref{masstoflux} shows the normalized mass-to-flux ratio, $\mu$, as a function of core mass, $M_{\rm prog}$. We find that all of our unstable cores are supercritical, which is to be expected, as we selected for unstable cores associated with recently formed star particles.
  
   \begin{figure}
   \centering
   \includegraphics[width=\hsize]{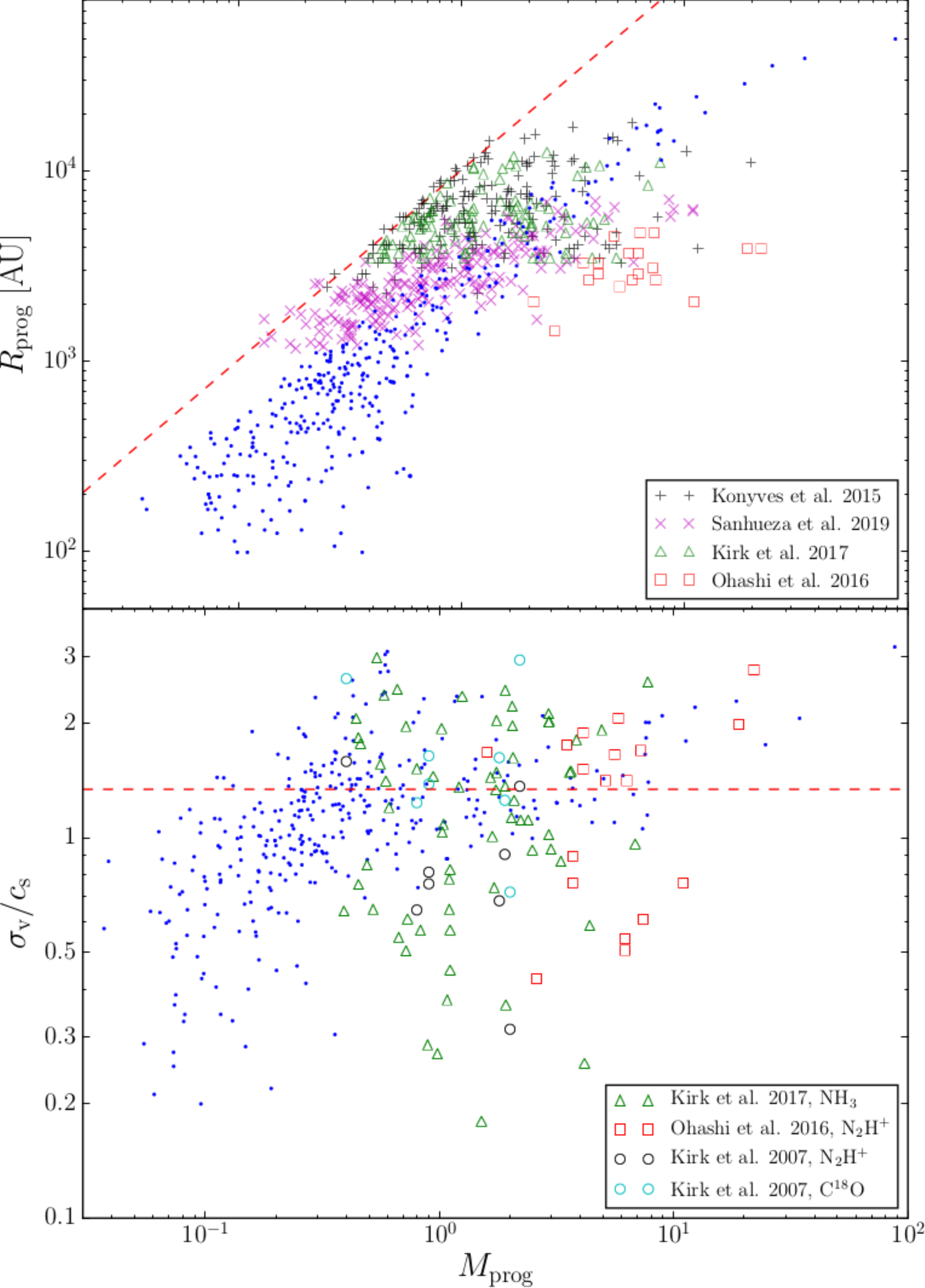}

      \caption{{\it Upper panel:} Progenitor core radius, $R_{\rm prog}$, as a function of the progenitor mass, $M_{\rm prog}$ (blue dots). All the cores are below the critical Bonnor-Ebert radius for a temperature of 10~K (dashed red line, $R_{\rm BE} = M_{\rm BE}/20.2$, where the mass is in $\rm M_\odot$ and the radius in parsecs). Observational radii and masses from several large surveys are also shown (various symbols, see the legend) for a qualitative comparison with the simulation (see Section~\ref{sectprog} for details). {\it Lower panel:} progenitor core rms sonic Mach number, $\sigma_{\rm v}/c_{\rm S}$, as a function of the progenitor mass, $M_{\rm prog}$ (blue dots). The red dashed line corresponds to the equipartition of kinetic and thermal energies ($E_{\rm k}/E_{\rm t} = 1$). Most of the progenitor cores are around energy equipartition, with $\mathcal{M_{\rm s}}$ rising slightly with increasing mass. The lower-mass cores ($M_{\rm prog} < 0.3 \; \rm M_\odot$) are predominately below the equipartition line and subsonic, with values as low as 0.2. Observational results from several large surveys are also shown (various symbols, see the legend) for comparison (see Section~\ref{sectprog} for details).
      }
         \label{mach}
   \end{figure}

To further describe the sample of progenitor cores, Figure ~\ref{mach} shows their size (upper panel) and their rms sonic Mach number (lower panel) as a function of their mass (blue dots). Although we make no attempt to retrieve observational properties of the cores through synthetic observations, it is still instructive to compare these quantities with some observational values. For that purpose, Figure ~\ref{mach} also shows values of sizes, rms Mach numbers and masses of prestellar cores from a number of large observational surveys.

The cores from the simulation are studied at the time when they have just started to collapse in order to identify the maximum value of their mass in the prestellar phase. Thus, our cores are in the transition from prestellar to protostellar, and they could also be compared with very young protostellar cores whose central star is still a small fraction of the core mass. For simplicity, we consider only observed cores in the prestellar phase.  

The gravitational stability of cores is often defined by reference to a critical Bonnor-Ebert sphere. The critical radius, $R_{\rm BE}$, of a Bonnor-Ebert sphere with a temperature of 10~K is given by $R_{\rm BE} = M_{\rm BE}/20.2$, where $M_{\rm BE}$ is the mass of the sphere in $\rm M_\odot$, and the radius is in parsecs. This expression for the critical radius as a function of the mass is shown by the red dashed line in the upper panel of Figure ~\ref{mach}. The progenitor cores from our simulation span a range of values between approximately 100~AU and 0.2~pc, with a median value of approximately 800~AU. They are all well below the critical Bonnor-Ebert radius, because they are selected as gravitationally unstable cores accounting also for their internal kinetic energy, as shown in Equation~(\ref{eqalphavire}).

For comparison, we show the observational values of the prestellar cores in Aquila from the Herschel Gould Belt Survey \citep{Konyves+15}, in Orion from a sub-sample of a SCUBA survey that includes followup line observations \citep[][NH$_3$ observations from \citet{Friesen+17}]{Kirk+17}, in a number of infrared dark clouds from a large ALMA survey \citep{Sanhueza+19}, and in another set of infrared dark clouds observed with ALMA where core velocity dispersions are also published \citep{Ohashi+16}. There is a significant overlap between the observed prestellar cores and the cores from our simulation, except that core sizes below 1,000~AU are missing in the observational samples, due to their spatial-resolution limit\footnote{
Although the least massive of our cores have the shortest free-fall times (they must be very dense to be gravitationally unstable), and so they may be more appropriately compared with cores in their initial protostellar phase, as mentioned above, in the surveys considered here even protostellar cores are all larger than approximately 1,000~AU, due to the resolution of the observations.}. Furthermore, the observed core masses at any given radius tend to extend to larger values than in our sample, probably also as an effect of the limited spatial resolution \citep[see discussion in \S~9.2 of][]{Padoan+20SN_massive}. Finally, for the Aquila and Orion cores we have selected all prestellar cores with a radius smaller than or equal to the critical Bonnor-Ebert radius for their mass, which explains the large number of cores near the critical radius line. Although such cores, and even those with half a critical Bonnor-Ebert mass, are usually considered prestellar cores in dust-continuum studies, a fraction of them are probably not gravitationally bound, due to their internal kinetic energy. 

To illustrate the importance of internal motions in the progenitor cores, the lower panel of Figure~\ref{mach} shows the one-dimensional velocity dispersion inside the cores, $\sigma_{\rm v}$, in units of the isothermal speed of sound for a temperature of 10~K, $c_{\rm S}\approx 0.18$~km~s$^{-1}$, as a function of the core mass, $M_{\rm prog}$. In the case of the observational samples, the sound speed value estimated in the corresponding papers is adopted. Most of the progenitor cores (blue dots) are within a factor of two from the equipartition between kinetic and thermal energy, marked by the horizontal dashed line. They show a trend of increasing rms Mach number with increasing core mass, and some of the least massive cores have an rms Mach number several times smaller than the value corresponding to the energy equipartition, or velocity dispersions as low as 1/5 of the sound speed. 

Apart from the prestellar cores in Orion and in the infrared dark clouds already shown in the upper panel of Figure~\ref{mach}, we consider also the velocity dispersion of prestellar cores in Perseus from \citet{Kirk+07}. The observed values overlap with those of the progenitor cores from our simulation. However, in the range of core masses between approximately 1 and 10~M$_{\odot}$, the observed cores extend to lower Mach number values than our progenitor cores. This partial discrepancy may have two origins. First, as mentioned above and extensively documented in \S~9.2 of \citet{Padoan+20SN_massive}, because of the limited spatial resolution of the observational surveys core masses may be systematically overestimated by large factors. If the observed cores were split into their lower-mass components, they may overlap in both velocity dispersion and mass with our low-mass progenitor cores in Figure~\ref{mach}. The second reason for the partial discrepancy is that we have computed the core rms Mach number from the ratio of kinetic and thermal energies, meaning that the rms velocity is density weighted. In the observations, the linewidths may more closely correspond to a velocity dispersion weighted by the square of the density, in the case of an optically-thin line like N$_2$H$^+$, so it is more strongly skewed towards the densest gas in the cores, where the velocity dispersion is usually smaller than in the outer region of the core. In the case of optically-thick lines, the velocity dispersion would be more representative of the outer layers of the cores, as illustrated by the comparison between the N$_2$H$^+$ and C$^{18}$O linewidths of the same cores in the survey by \citet{Kirk+07}. For each core, the velocity dispersion based on the C$^{18}$O linewidth (cyan circles) is always larger than that based on the N$_2$H$^+$ line (black circles). As a result, the C$^{18}$O rms Mach numbers cover a very similar range of values as that of our progenitor cores. 

As mentioned above, we make no attempt here to derive observational core properties with synthetic observations, as this work is primarily focused on testing a fundamental theoretical assumption of star-formation models. This comparison with observed prestellar cores is shown to illustrate that the values of the physical parameters of the cores from our simulation are reasonable, which does not require that such values cover the full range of parameter space from the observations. All the physical parameters of the progenitor cores used in this study are listed in Table~\ref{tableprog}. The Table, as well as other supplemental material, can also be obtained from a dedicated public URL (http://www.erda.dk/vgrid/core-mass-function/).

\begin{table*}
\caption{Stellar and progenitor parameters for a set of 10 stars and their progenitors. The full table is included as an electronic download. The columns are: 1 star index, 2 snapshot number when the new star was recorded, 3 final mass of the star, 4 mass of the progenitor, 5 radius of the progenitor (as an equivalent volume sphere), 6 radius within which the star accretes 95\% of its mass, 7 one-dimensional sonic Mach number, 8 kinetic-to-gravitational energy ratio, 9 thermal-to-gravitational energy ratio, 10 inverse of normalized mass-to-flux ratio, 11 final accretion rate as a fraction of the final stellar mass, 12 fraction of progenitor mass which is accreted by other stars, 13 fraction of progenitor mass which is accreted by the star itself, 14 fraction of final stellar mass already present in the progenitor, and 15 flag for the state of the progenitor (1 for single star, 2 for multiple stars born at the same time). The fraction of progenitor mass which remains unaccreted by any star can be derived from $f_{\rm tr,unacc} = 1 - (f_{\rm tr,other} + f_{\rm tr,prog}$).}             
\label{tableprog}      
\centering                          
\begin{tabular}{c c c c c c c c c c c c c c c }        
\hline\hline                 
Star & Snap & $M_{\rm star}$ & $M_{\rm prog}$ & $R_{\rm prog}$ & $R_{95}$ & $\sigma_{\rm v}/c_{\rm s}$ & $E_{\rm k}/E_{\rm g}$ & $E_{\rm t}/E_{\rm g}$ & $1/\mu$ & $\dot{M} / M_{\rm star}$ & $f_{\rm tr,other}$ & $f_{\rm tr,prog}$ & $f_{\rm tr,star}$ & Flag\\    
 & & $[M_\odot]$ & $[M_\odot]$ & $[\rm AU]$ & $[\rm AU]$ & & & & & $[\rm Myr^{-1}]$ & [\%] & [\%] & [\%] & \\ 
\hline                        
... \\
41 & 73 &  0.13 &  0.26 &    451 &   1584 & 1.18 & 0.22 & 0.29 & 0.09 & 0.00e+00 & 16.1 & 83.9 & 85.7 & 1 \\
42 & 73 &  0.69 & 24.65 &  35913 &   9678 & 1.76 & 0.41 & 0.24 & 0.07 & 0.00e+00 & 32.4 &  5.4 & 97.4 & 1 \\
43 & 73 &  0.61 &  0.49 &    778 &   5522 & 2.25 & 0.74 & 0.26 & 0.09 & 0.00e+00 &  0.0 & 100.0 & 39.7 & 1 \\
44 & 73 &  0.12 &  0.13 &    325 &  46197 & 1.37 & 0.42 & 0.40 & 0.02 & 7.50e-03 & 29.8 & 70.2 & 37.6 & 1 \\
45 & 73 &  0.11 &  0.59 &    248 &   5275 & 3.03 & 0.35 & 0.07 & 0.03 & 0.00e+00 & 82.7 & 17.3 & 48.0 & 2 \\
46 & 73 &  0.57 &  0.59 &    251 &  40109 & 2.85 & 0.32 & 0.07 & 0.03 & 0.00e+00 & 82.2 & 17.8 &  9.2 & 2 \\
47 & 73 &  0.05 &  0.12 &    337 &   7163 & 1.44 & 0.53 & 0.46 & 0.09 & 6.33e-02 & 48.9 & 51.1 & 60.2 & 1 \\
48 & 73 &  0.15 &  0.11 &    251 &   9515 & 1.12 & 0.26 & 0.37 & 0.09 & 0.00e+00 & 13.7 & 86.3 & 31.4 & 1 \\
49 & 74 &  1.19 &  0.43 &    769 & 117127 & 2.07 & 0.70 & 0.29 & 0.07 & 3.20e-01 &  0.2 & 99.8 & 17.9 & 1 \\
50 & 74 &  0.15 &  0.16 &    293 &   4358 & 0.55 & 0.05 & 0.29 & 0.02 & 0.00e+00 &  0.8 & 99.2 & 53.8 & 1 \\
... \\

\hline                                   
\end{tabular}
\end{table*}

   \begin{figure}
   \centering
   \includegraphics[width=\hsize]{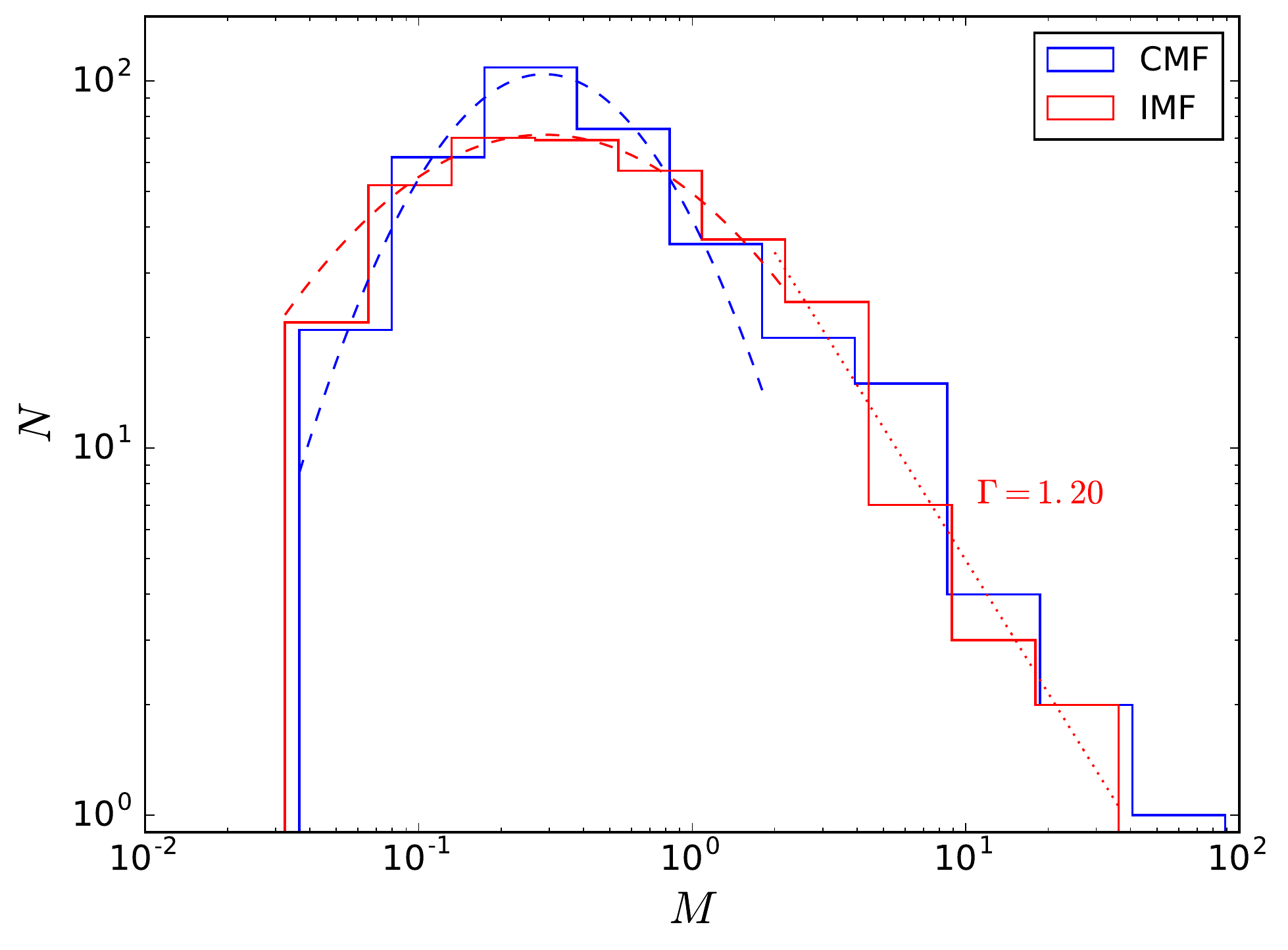}
      \caption{Histogram of prestellar progenitor masses (blue, 344 cores) and the final masses of the stars born in them (red, 344 stars), with bin sizes 0.34 and 0.30 dex, respectively. Dashed lines are lognormal fits ($N \propto \exp(-(\log_{10} M - \log_{10} M_{\rm peak})^2 / 2 \sigma^2$) to masses lower than $2 M_\odot$ (CMF: $M_{\rm peak}=0.29 \; M_\odot$, $\sigma=0.40$; IMF: $M_{\rm peak}=0.29 \; M_\odot$, $\sigma=0.63$) and the red dotted line is a power-law fit ($N \propto M^{-\Gamma}$) to stellar masses above $2 M_\odot$ with $\Gamma = 1.20$. 
              }
         \label{masshist}
   \end{figure}

\section{Progenitor Masses versus Stellar Masses}
\label{progvssink}

\subsection{Statistical comparison}
\label{statistical}

The main goal of this work is to test the hypothesis of the core-collapse model, which we express as $M_{\rm star}\approx\epsilon_{\rm prog} M_{\rm prog}$, with $\epsilon_{\rm prog}\lesssim 0.5$ and approximately independent of mass, using previously defined quantities. Before comparing cores and stars one-to-one, a look at their mass distribution is already instructive. Figure~\ref{masshist} shows the mass distribution of the progenitors selected at the star-formation snapshots and that of the stars at the end of the simulation. Only the 344 prestellar progenitors, those without older star particles in them (see Sect.~\ref{sectprog}), are included in the figure. Both the progenitor CMF and the IMF distribution peak at $\sim 0.3 \rm M_\odot$. Our resolution convergence test in Appendix~\ref{app_parameters} indicates that the mass peak of the CMF converges towards a well-defined value $M_{\rm conv} \simeq 0.26 M_\odot$, almost identical to the value of the IMF peak. This is problematic for the core-collapse model, as it would imply $\epsilon_{\rm prog}\approx 1$.

\begin{figure}
   \centering
   \includegraphics[width=\hsize]{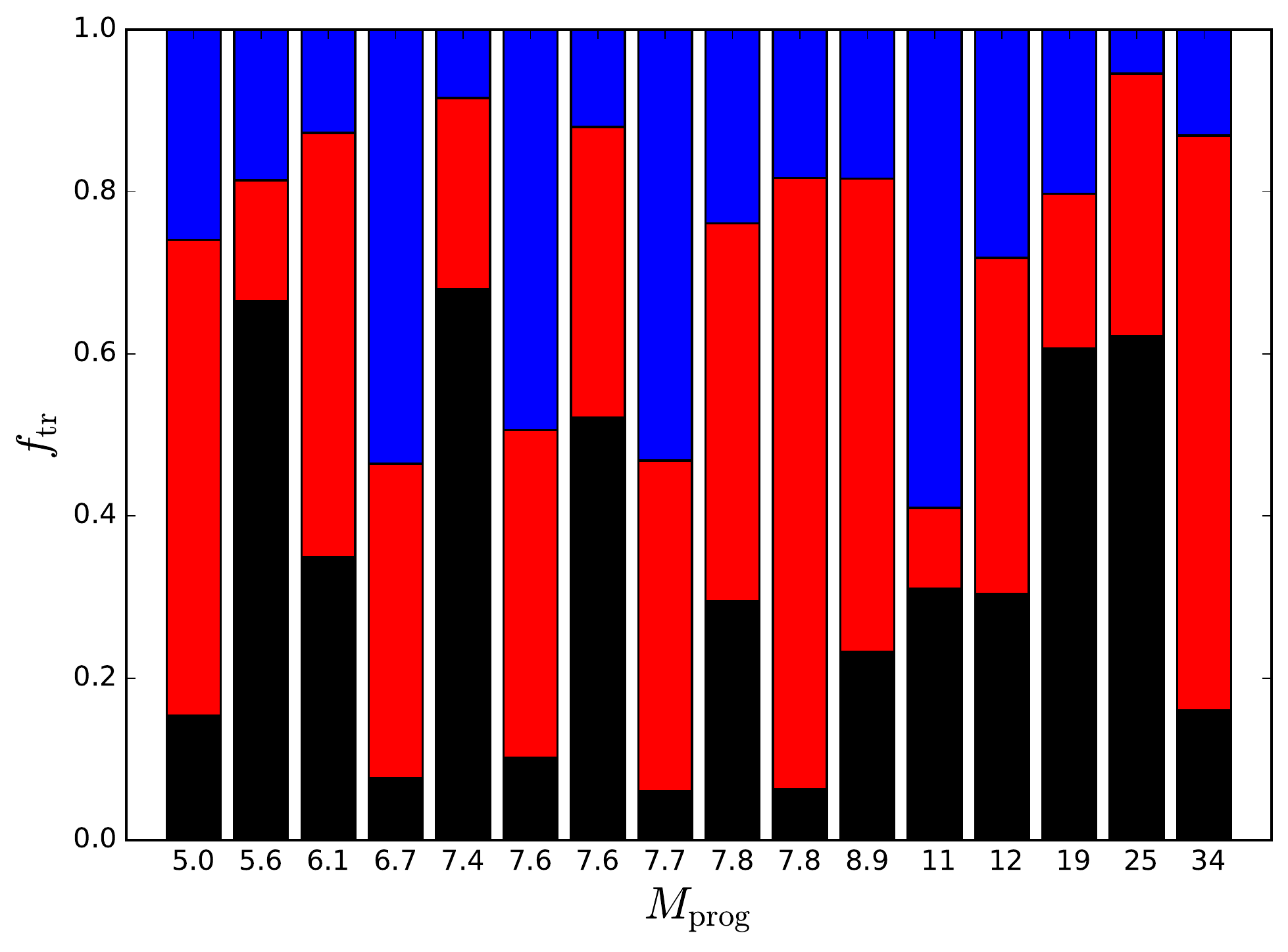}
      \caption{Bar chart of the fractions where the progenitor mass goes, for progenitors with $M_{\rm prog} > 5 M_\odot$ (the most massive cores in our sample). The total mass of the progenitor is stated on the x-axis, and the fractions are unaccreted mass ($f_{\rm tr,unacc}$, black), mass accreted by other stars ($f_{\rm tr,other}$, red), and mass accreted by the star whose progenitor the core is ($f_{\rm tr,prog}$, blue). Two high-mass cores, $7.5$ and $88 \; M_\odot$, are not shown as they are identified less than 100~kyr before the end of the simulation. 
              }
         \label{ftr_massive}
   \end{figure}

The similarity of the mass distributions continues to the high-mass tail, as well. However, this should not be taken to mean that the high-mass progenitors are undergoing a monolithic collapse. In order to show this, we used the tracer particles (see \S~\ref{sectprog}) to study how the gas from the progenitor core is either accreted by the primary star ($f_{\rm tr,prog}$, blue), by other stars ($f_{\rm tr,other}$, red), or remains unaccreted ($f_{\rm tr,unacc}$, black). These mass fractions, defined in \S~\ref{sectprog}, are shown in Figure~\ref{ftr_massive} as a bar chart for the 16 cores with masses larger than 5~$M_\odot$, where the x-axis is ordered by growing progenitor mass. The mass fractions for two high-mass cores, $7.5$ and $88 \; M_\odot$, are not shown as they are identified less than 100~kyr before the end of the simulation, and most of their mass is unaccreted. The bar charts show that in the majority of the massive progenitors, the primary star only accretes less than half of the progenitor's mass, often much less. 

\begin{figure}
   \centering
   \includegraphics[width=\hsize]{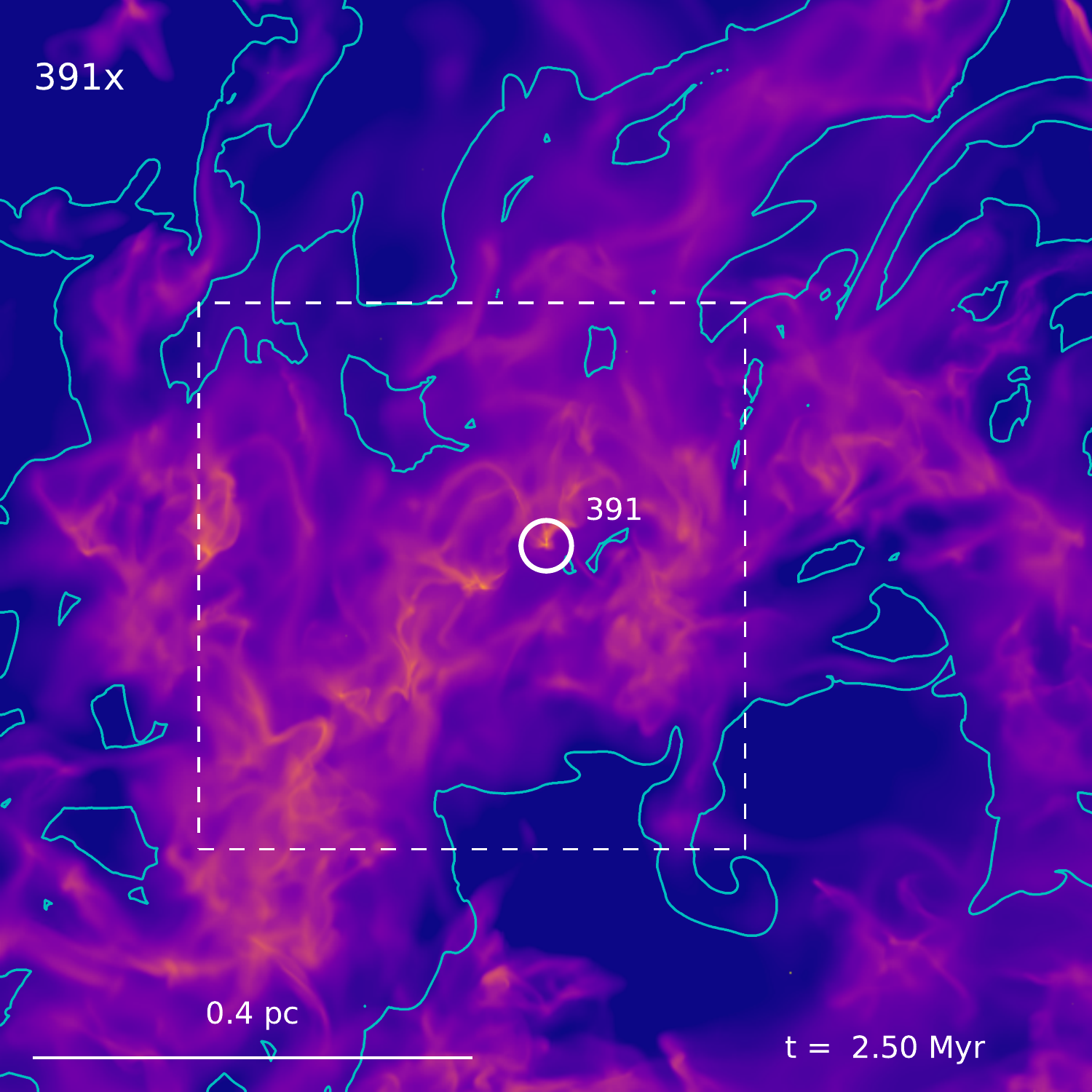}
   \includegraphics[width=\hsize]{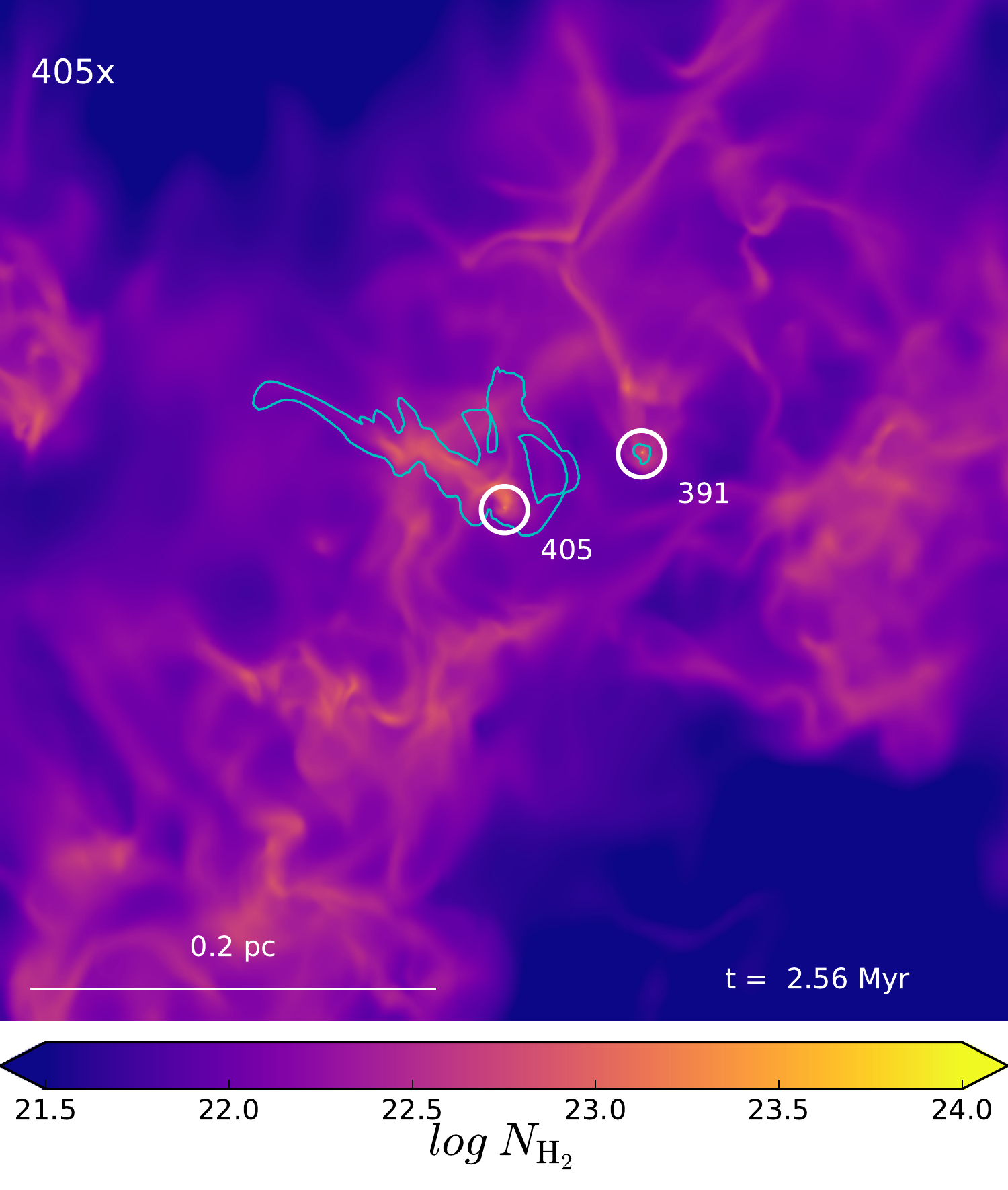}   
      \caption{{\it Upper panel:} Column density map of the progenitor core of star 391, seen from x direction. Cyan contours are drawn where there is at least one cell belonging to the progenitor core on the line of sight. The star being formed is in the middle (white circle and number), but it is clear that there are other nodes forming in this high-mass ($88 \; M_\odot$) core, too. The white dashed box shows the approximate location of the lower panel. {\it Lower panel:} As above, but for star 405, 66~kyr (3 snapshots) later. The density structure around star 391 is clearly identifiable, but the bound core around it is just a protostellar envelope with only $0.02 \; M_\odot$ (the protostar itself is $0.14 \; M_\odot$ at this time). The progenitor core associated with star 405 is $1.1 \; M_\odot$. 
              }
         \label{NH_391x}
   \end{figure}

This result is also illustrated by the fact that the massive cores show a lot of internal structure, and their time evolution always results in their fragmentation. This is exemplified for star 391 in Figure~\ref{NH_391x}, where the lower panel shows that another star has formed 66~kyr later and the bound core around star 391 has shrunk to a mere $0.02 \; M_\odot$ protostellar envelope (the protostar itself is $0.14 \; M_\odot$). The vast majority of the gas in its $88 \; M_\odot$ progenitor core is no longer bound to star 391. Even accounting for the protostellar mass, the bound core around star 391 is not even the locally dominant core, as the mass of the progenitor core of star 405 is $1.1 \; M_\odot$.

Thus, the statistical similarity between the high-mass tails of the CMF and the stellar IMF does not imply a monolithic collapse of the massive cores as further discussed in the next subsection.
   
\subsection{One-to-one comparison}
\label{one-to-one}

\begin{figure}
   \centering
   \includegraphics[width=\hsize]{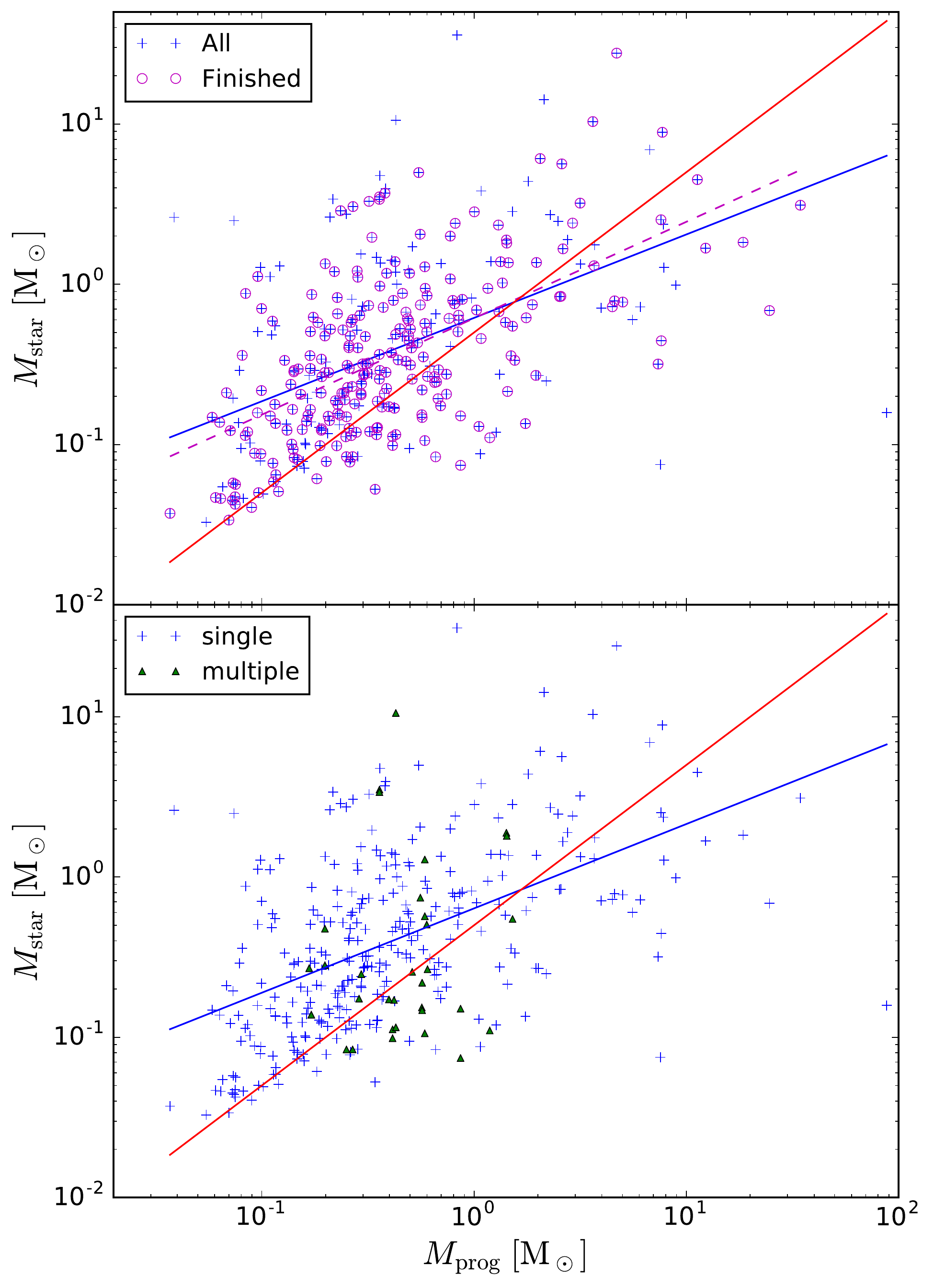}
      \caption{Scatter plots of progenitor mass, $M_{\rm prog}$, and final stellar mass, $M_{\rm star}$. 
      {\it Upper panel}: Blue crosses are all the stars with detected progenitors, and magenta circles are the stars that have also stopped accreting by the end of the simulation (needing 10 Myr or longer to double their mass at the current accretion rate). The blue line is a power-law fit ($M_{\rm star} \propto M_{\rm prog}^{a}$) to all stars ($a=0.52)$, while the magenta dashed line is a power-law fit to the stars that have stopped accreting ($a=0.60$). Pearson's correlation coefficient, $r$, for the blue and the magenta dashed line is 0.51 and 0.57, respectively. The red line shows the relation $M_{\rm star}=\epsilon_{\rm prog} M_{\rm prog}$, with $\epsilon_{\rm prog}=\epsilon_{\rm acc} = 0.5$, the value adopted in the simulation. 
      {\it Lower panel}: Same as the upper panel, but distinguishing progenitors with only one star (blue crosses) and progenitors with multiple stars born at the same time within them (green triangle). The blue line is a power-law fit to the single stars, with $a=0.53$ and $r=0.53$. 
              }
         \label{star_to_progenitor_mass_fit}
   \end{figure}

The one-to-one relation between progenitor masses and final stellar masses is addressed by the scatter plots in Figure~\ref{star_to_progenitor_mass_fit}. The figure shows that, for a given core mass, there is a scatter of about two orders of magnitude in the resulting stellar masses (Pearson's correlation coefficient, $r$, is $0.51$ for all stars). The top panel of Fig.~\ref{star_to_progenitor_mass_fit} distinguishes the stars that have finished accreting (magenta open circles). Limiting the sample to these stars does not make the correlation significantly stronger ($r = 0.57$). The correlation is even worse at lower resolution (see Appendix~\ref{app_parameters}).

The relatively weak correlation between core and stellar masses is in conflict with the core-collapse model, because it shows that even if we know the mass of the progenitor core at the birth time of the star, we cannot predict the final stellar mass with any reasonable accuracy. Furthermore, a least-square fit to the data points gives a slope of $a=0.52$ for all stars (blue line), and $a = 0.60$ for the stars that have finished accreting (magenta dashed line). This shows that even the average relation between core and stellar masses is inconsistent with the idea of a constant efficiency factor, $\epsilon_{\rm prog}$, as a constant efficiency would imply a slope $a=1$ (red solid line in Figure~\ref{star_to_progenitor_mass_fit}).

Some of the high-mass progenitors can be expected to fragment and contribute to several stars, as already mentioned in Sect.~\ref{statistical}. In the bottom panel of Fig.~\ref{star_to_progenitor_mass_fit} we distinguish between stars that formed alone or accompanied by other stars. Progenitors with multiple stars have an elevated progenitor mass on average at lower stellar masses. This is easily understood, as those progenitors are feeding two or more stars, while at higher masses, the progenitor alone is not enough for those massive stars to grow. If the star is the first one to form inside the core, even if it appears later in another progenitors, it is still counted as single (when we check for stars that appear later in other progenitors, we do not find a clear correlation with their own progenitor masses, so this does not bias the result). For single stars, the Pearson's correlation coefficient and the slope are $r=0.53$ and $a=0.53$, comparable to that of all the stars that have finished accreting. In the case of the stars that grow beyond their progenitor core mass, more mass needs to come in from outside the progenitor. Converging flows may continue feeding mass into the core from beyond its gravitationally-bound limit, allowing the star to accrete much more mass than what was initially available from the core. This is the case in particular with high-mass stars, as further documented in the following, using tracer particles.

   \begin{figure}
   \centering
   \includegraphics[width=\hsize]{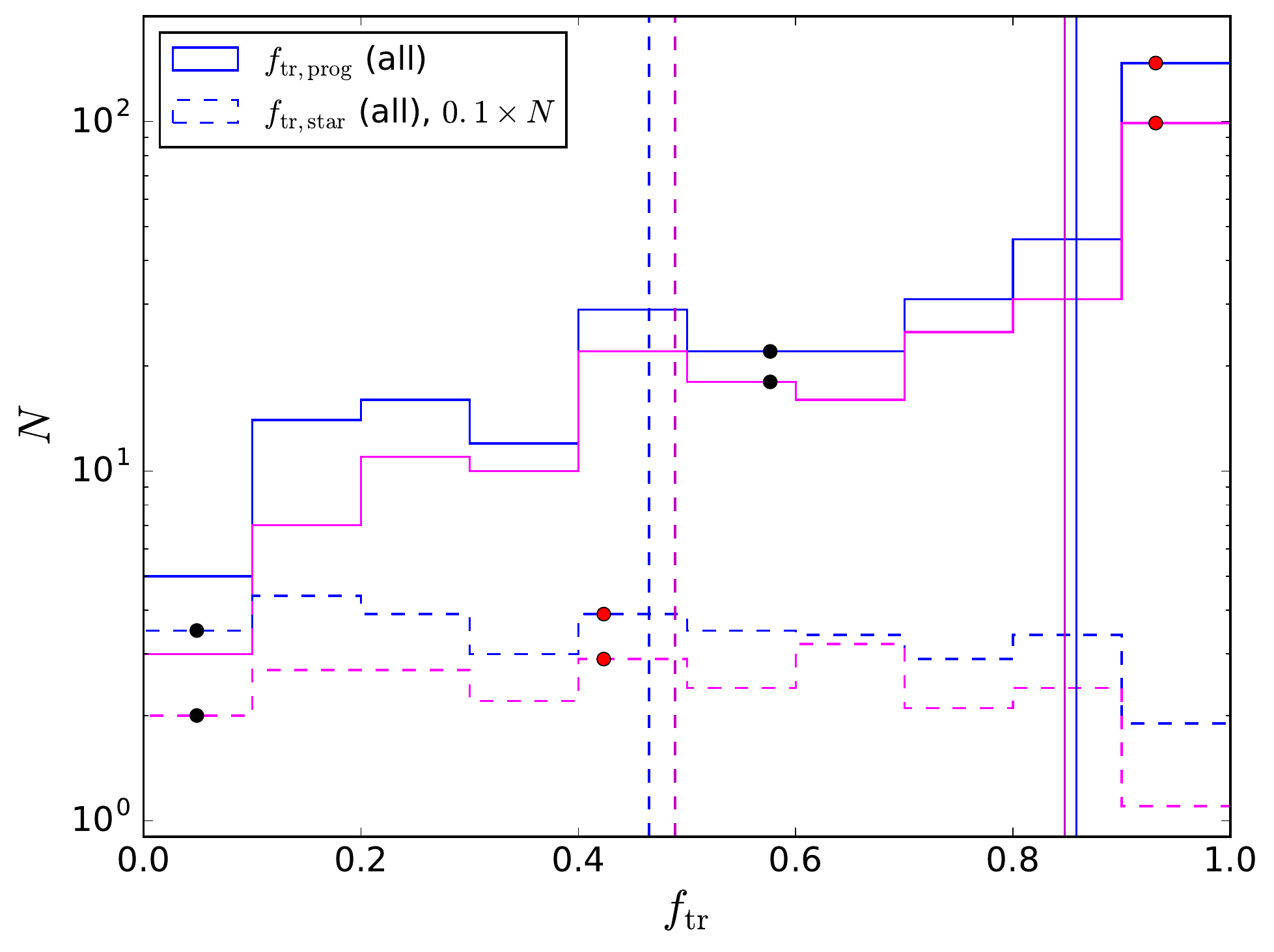}
      \caption{Histograms of progenitor mass fraction, $f_{\rm tr,prog}$, that is the fraction of the progenitor mass that is accreted on the star (solid lines), and stellar mass fraction, $f_{\rm tr,star}$, that is the fraction of the stellar mass that came from the progenitor (dashed lines). The dashed lines have been shifted down by a factor of 10, to avoid confusion with the solid lines. The blue lines are for all stars, while magenta lines are for the stars that have effectively finished accreting. Vertical lines mark the median value for each sample. Black (high-mass star) and red (low-mass star) dots show the fractions for the example progenitor cores in Figure~\ref{tracerexample}.
              }
         \label{tracerfraction}
   \end{figure}

\begin{figure}
   \centering
   \includegraphics[width=\hsize]{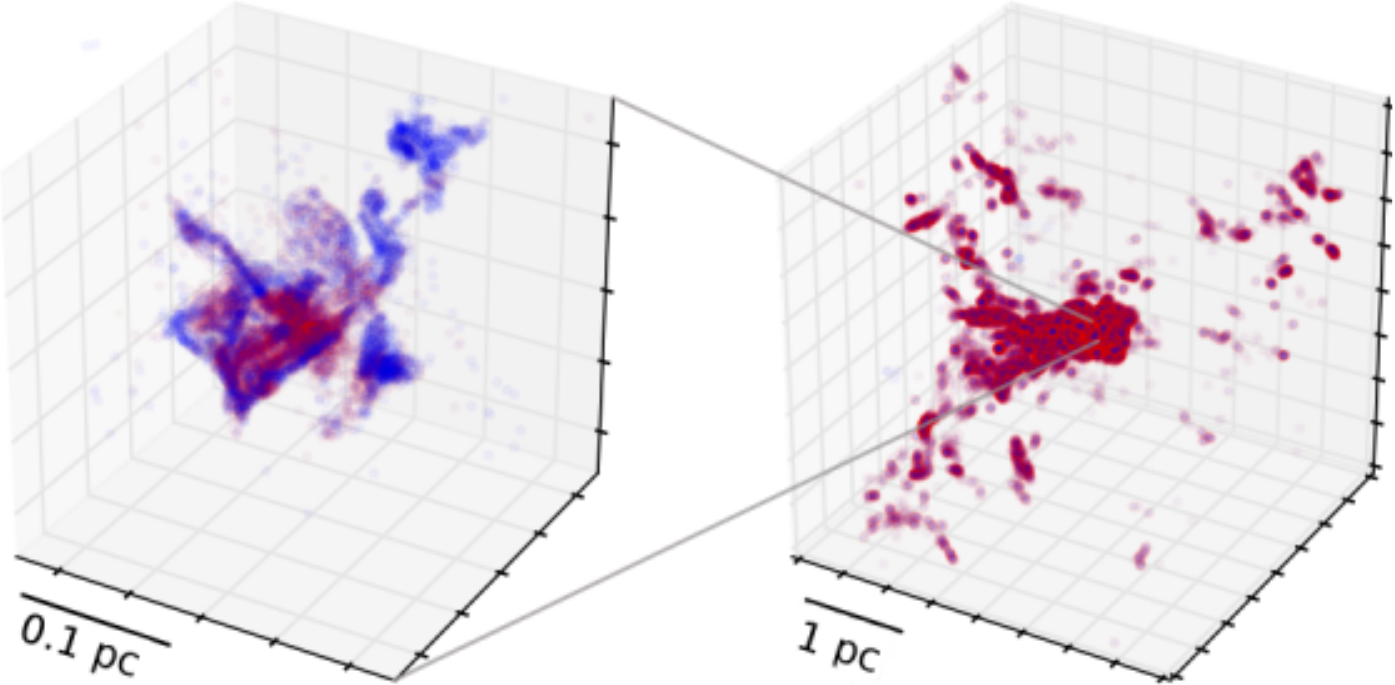}
   \includegraphics[width=\hsize]{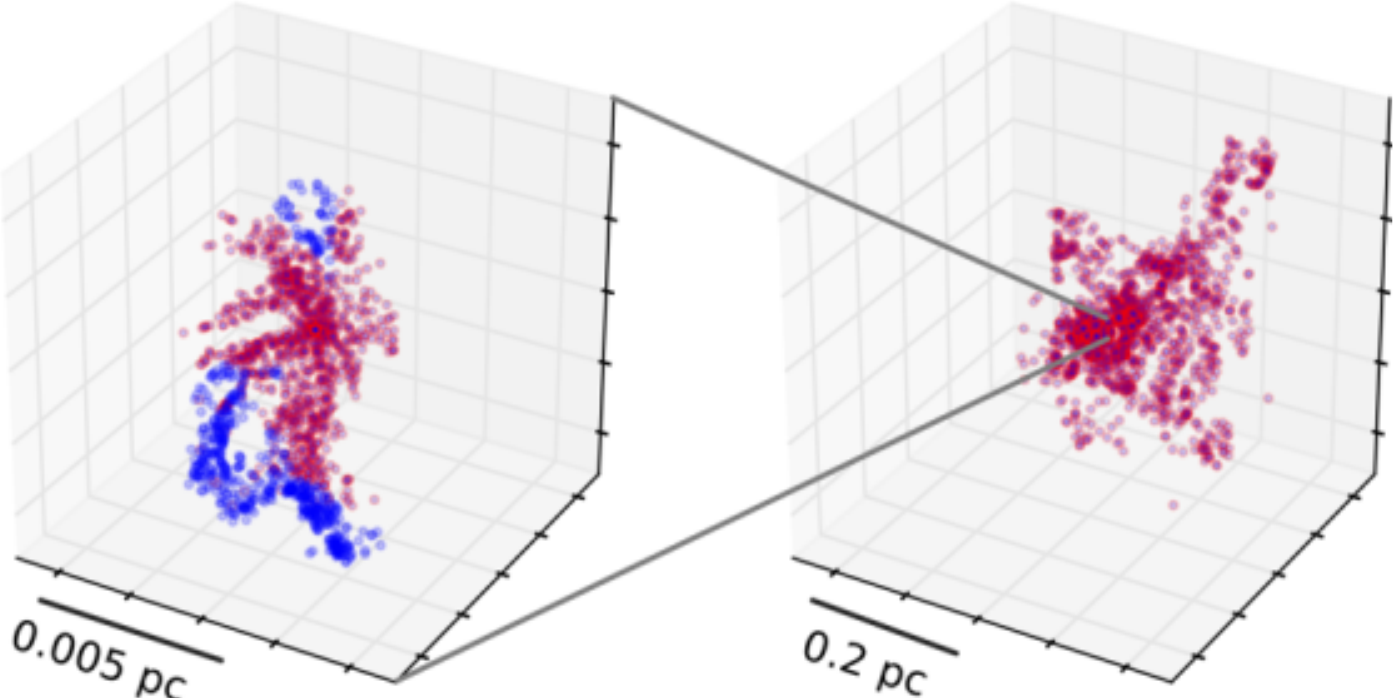}   
      \caption{Examples of progenitor cores that will form a high-mass star, $M_{\rm star} = 28$~M$_\odot$ (upper panels), and a low-mass star, $M_{\rm star} = 0.21$~M$_\odot$ (lower panels), visualized through their tracer particles. {\it Left panels}: The 0.25~pc (top) and 0.01~pc (bottom) central regions of the 0.5~pc subcubes where the two cores were selected. Red dots are tracers that will accrete onto the star, blue dots are the remaining tracers of the cores. {\it Right panels}: The whole 4~pc simulation box (top), and a 0.5~pc subcube (bottom) centred around the same star and at the same (star-formation) snapshot as the left panels. The red dots are the locations of all the tracers that will accrete onto the stars by the end of the simulation. A significant fraction, $\sim90\%$ (top) and $\sim60\%$ (bottom), of the final stellar mass is found outside the progenitor core.
              }
         \label{tracerexample}
   \end{figure}

Figure~\ref{tracerfraction} shows the histograms of mass ratios, $f_{\rm tr,prog}$ (solid lines) and $f_{\rm tr,star}$ (dashed lines). The first fraction, $f_{\rm tr,prog}$ is the fraction of the progenitor mass that is accreted onto the star; the second fraction, $f_{\rm tr,star}$, is the fraction of the final stellar mass that originates from the progenitor core, as explained in Section~\ref{sectprog}. Black lines are for all stars, while magenta lines are for stars that have stopped accreting, as explained previously. Vertical lines show the median values. The dots are the values of these fractions for the example progenitors depicted in Figure~\ref{tracerexample}, black for high-mass and red for low-mass. Most stars have a progenitor mass fraction, $f_{\rm tr,prog}$, higher than 0.5, and the distribution of $f_{\rm tr,prog}$ has a strong peak at $f_{\rm tr,prog}=0.9-1.0$. The median value is $\approx 0.85$, meaning that for half of the progenitors, 85\% or more of the progenitor mass eventually accretes onto the star. However, the stellar mass fraction, $f_{\rm tr,star}$, has an almost uniform distribution (dashed lines), with a median value of $\approx 0.5$. Thus, many stars have to accrete mass from outside their initial progenitor. 

\begin{figure}
   \centering
   \includegraphics[width=\hsize]{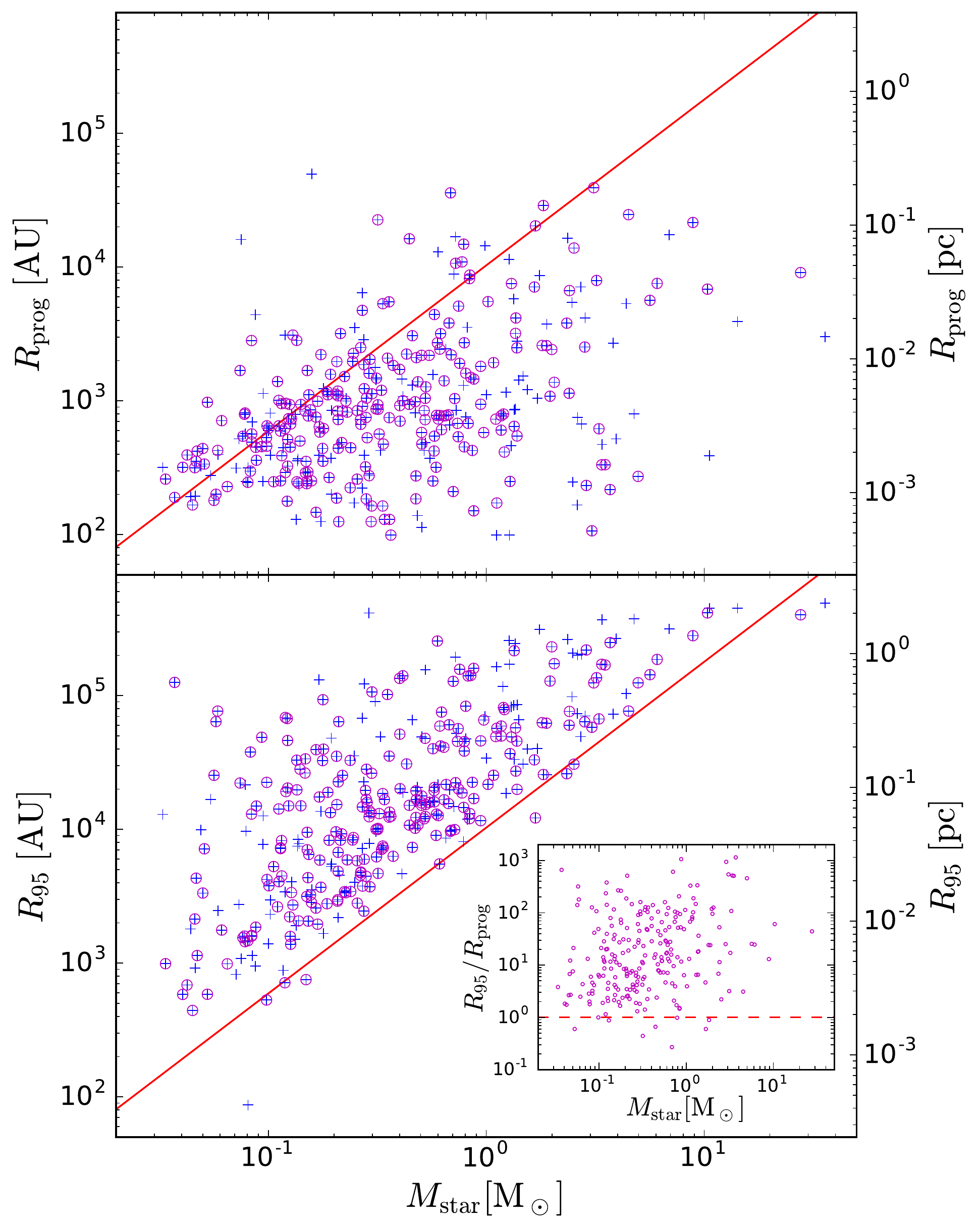}
      \caption{Scatter plots of the progenitor radius, $R_{\rm prog}$ (upper panel), and the inflow radius, $R_{\rm 95}$ (lower panel), as a function of the final stellar mass, $M_{\rm star}$. The left-hand y-axis is in AU units, while the right-hand is in pc units. Since the simulation box size is 4~pc, centred on each star for this analysis, and randomly driven on scales between the box size and half the box size, on scales above 1~pc $R_{\rm 95}$ may be influenced by the limited box size. Blue crosses are for all stars, and magenta circles signal stars that have finished accreting. The red line is the same in both panels, $R_{\rm 95} = 0.05 \rm \; pc \times (M_{\rm star} / 1 \; \rm M_\odot)^{1.24}$, from Figure~23 of \citet{Padoan+20SN_massive}. The inset in the lower panel shows the ratio of $R_{\rm 95}/R_{\rm prog}$ as a function of the final stellar mass for stars that have finished accreting, with the dashed red line indicating a ratio of one. There is essentially no correlation of this ratio with the final stellar mass.
              }
         \label{R95}
   \end{figure}

The origin of the mass that feeds the growth of a star is further illustrated in Figure~\ref{tracerexample} and Figure~\ref{R95}. Fig.~\ref{tracerexample} shows two example progenitors, where $\sim 95\%$ of the final stellar mass is accreted from the entire 4~pc simulation box for the high-mass example, and close to 60\% from the 0.5~pc subcube for the low-mass example. Fig.~\ref{R95} shows scatter plots of the progenitor radius, $R_{\rm prog}$ (upper panel), and the inflow radius, $R_{\rm 95}$ (lower panel), as a function of the final stellar mass, $M_{\rm star}$, for all 344 stars. Values of $R_{\rm 95}$ above 1~pc may be affected by the finite size of the simulation box, 4~pc, centred on each star for this analysis. The box size limits the distance where tracer particles can originate from, and the random driving force, applied to scales between the box size and half the box size may also affect the particle trajectories at those scales. The red line is the same in both panels, $R_{\rm 95} = 0.05 \rm \; pc \times (M_{\rm star} / 1 \; \rm M_\odot)^{1.24}$, taken from Figure~23 of \citet{Padoan+20SN_massive}, where it was obtained as a fit to the inflow radius of the stars with the highest accretion rate and was shown to be a good approximation to the lower envelope of the $R_{\rm 95}$ versus $M_{\rm star}$ plot. In \citet{Padoan+20SN_massive} that plot covered stellar masses above 2-3~M$_{\odot}$. Here, the same function is found to be a good approximation to the lower envelope of the plot for all stellar masses, down to brown dwarfs. 

The inset in the lower panel shows the ratio of $R_{\rm 95}/R_{\rm prog}$ as a function of the final stellar mass for stars that have already finished accreting, with the dashed red line indicating a ratio of unity. The scatter plot covers rather uniformly a range of values of $R_{\rm 95}/R_{\rm prog}$ between approximately 1 and $10^3$, with a median of 14. There is no correlation between the ratio and the final stellar mass, which indicates that even lower-mass stars are influenced by proportionally as large a region around them as the more massive stars. Of the five lowest ratio cases (with a ratio < 1), four are massive progenitors ($M_{\rm prog} > 7 \; \rm M_\odot$) that later undergo sub-fragmentation to form relatively low-mass stars (0.3, 0.4, 0.7 and 1.7~$\rm M_\odot$), and the remaining one is a low-mass progenitor ($M_{\rm prog}=0.3 \; \rm M_\odot$) that results in the formation of a brown dwarf (0.05~$\rm M_\odot$).

\begin{figure}
   \centering
   \includegraphics[width=\hsize]{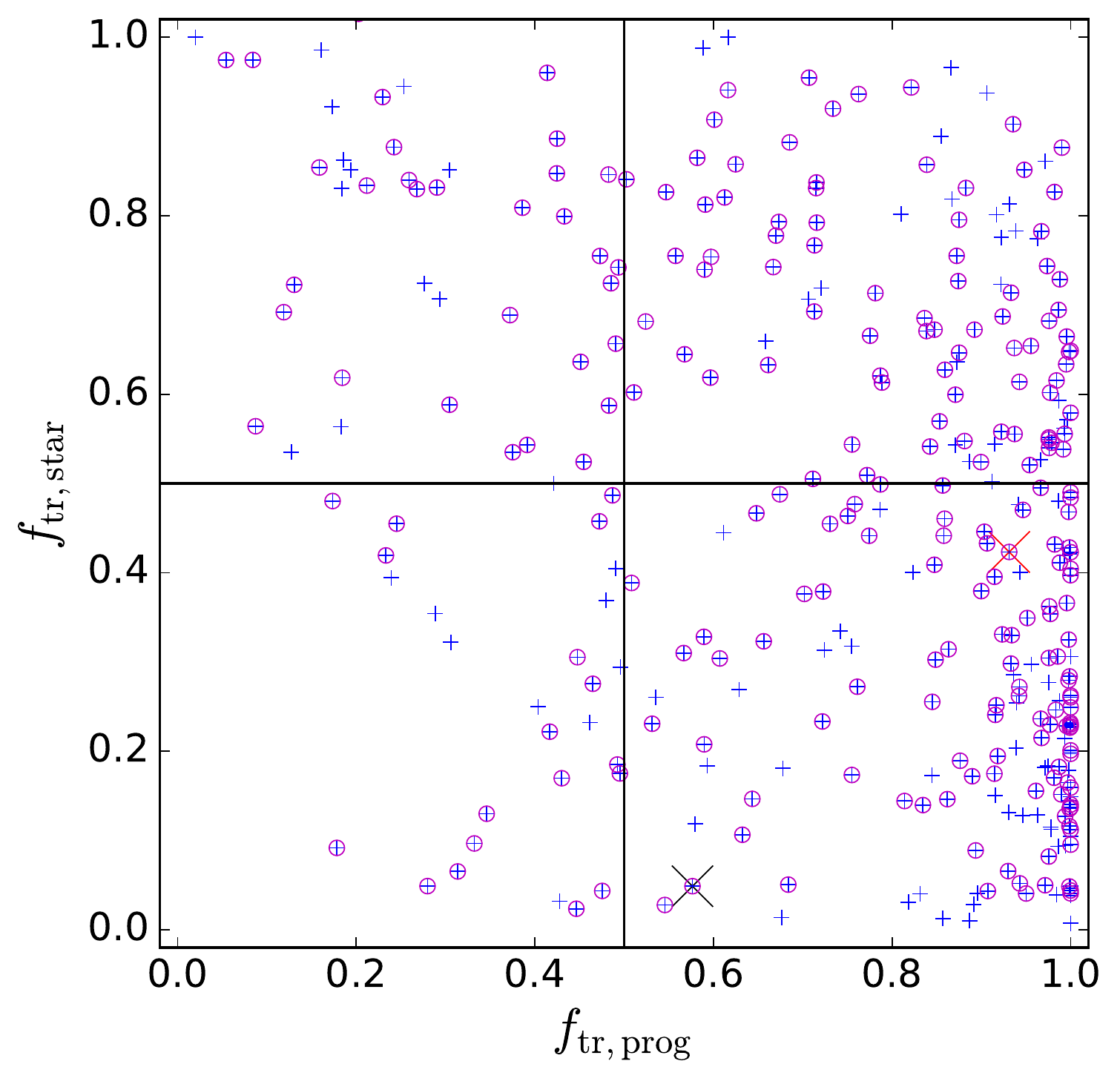}
      \caption{Scatter plot of progenitor mass fraction, $f_{\rm tr,prog}$ (the fraction of the progenitor mass that is accreted on the star), and stellar mass fraction, $f_{\rm tr,star}$ (the fraction of the stellar mass that came from the progenitor). The blue pluses are for all stars, while magenta open circles are for the stars that have effectively finished accreting. Black and red crosses shows the fractions for the example progenitor cores in Figure~\ref{tracerexample}, high-mass and low-mass respectively. Only $\sim35\%$ of the 344 star-progenitor pairs are in the upper right quadrant, where both $f_{\rm tr,prog}$ and $f_{\rm tr,star}$ are higher than 0.5.
              }
         \label{tracer_scatter}
   \end{figure}

Figure~\ref{tracer_scatter} shows the progenitor mass fraction, $f_{\rm tr,prog}$, and the stellar mass fraction, $f_{\rm tr,star}$, in a scatter plot. The ratio of the stellar mass and the progenitor mass is $M_{\rm star}/M_{\rm prog}\propto f_{\rm tr,prog}/f_{\rm tr,star}$, so the core-collapse model may appear to be satisfied along the diagonal of Figure~\ref{tracer_scatter}, where $f_{\rm tr,prog}=f_{\rm tr,star}$, or, equivalently, the red line in Figure~\ref{star_to_progenitor_mass_fit}. However, for many stars where $M_{\rm star}\sim \epsilon_{\rm prog} M_{\rm prog}$, this agreement is only due to the fact that both $f_{\rm tr,prog}$ and $f_{\rm tr,star}$ are $<1$, so both mass fractions actually violate the physical assumption of the core collapse model (although not its mathematical expression based only on $M_{\rm star}$ and $M_{\rm prog}$). In other words, the core-collapse model is truly satisfied, within an error of a factor of two, only in the upper right quadrant of the scatter plot in Figure~\ref{tracer_scatter}. Only $\sim35\%$ of the 344 stars are in the upper right quadrant, where both $f_{\rm tr,prog}$ and $f_{\rm tr,star}$ are higher than 0.5. Thus, from the perspective of the tracer particles, the one-to-one relation between core and stellar masses assumed by the core-collapse model would be correct within a factor of two for only a minority of the cores.

\section{Discussion}

\subsection{The Core-Collapse Model}

We have compared the final stellar masses with those of their prestellar cores. Accurate predictions of both masses are difficult to obtain, due to the complexity of the star-formation process. However, we argue that uncertainties related to such mass determinations are not critical to our approach to test the core-collapse model and do not affect our conclusions.  The mass of a star at the end of the simulation is a good approximation to the final stellar mass, particularly for the stars that were tagged as finished accreting. Because the simulation does not resolve jets and outflows, only half of the accreting mass is assigned to the star particle, a typical value found in theoretical models \citep[e.g.][]{Matzner+McKee2000}. This simple approach to modeling the mass loss from protostellar jets and outflows introduces some uncertainty in the final stellar mass. However, this uncertainty does not affect the fundamental question addressed here, which is the origin of the stellar mass reservoir, irrespective of how much of that may get ejected during its formation.

The estimation of the prestellar core mass is particularly difficult, because they arise within dense filaments as a result of converging flows in the turbulent gas, so they rarely appear as isolated objects with simple morphology. Furthermore, a comparison of prestellar core masses from a simulation with those from observations would have to address a number of observational limitations that are briefly mentioned in \S~\ref{sect_obs}. This would require an analysis of the simulation based on synthetic observations, and a method of core selection following the observational procedures. We have not attempted that, as our main focus is to test a theoretical idea. In the theoretical models, the stellar progenitors are well-defined entities. In the IMF models of \citet{Hennebelle+Chabrier08imf} and \citet{Hopkins12imf} the stellar-mass reservoir is a gravitationally bound, overdense region in the turbulent flow, consistent with the core-collapse scenario. Such mass reservoir may not correspond precisely to what the observers identify as prestellar cores (e.g. it may have to contract somewhat before reaching a characteristic core density), or with analytical models of isolated cores \citep[e.g.][]{McKee+Tan02,McKee+Tan03}, as it is selected through a statistical description of a complex turbulent flow. However, that complexity is fully accounted for in our simulation, and our progenitor cores are defined as bound regions, as in the models. So the progenitor masses we derive are relevant for testing the IMF models based on the idea of core-collapse, and our approach is not affected by the uncertainties related to a comparison with the observations.

   \begin{figure}
   \centering
   \includegraphics[width=\hsize]{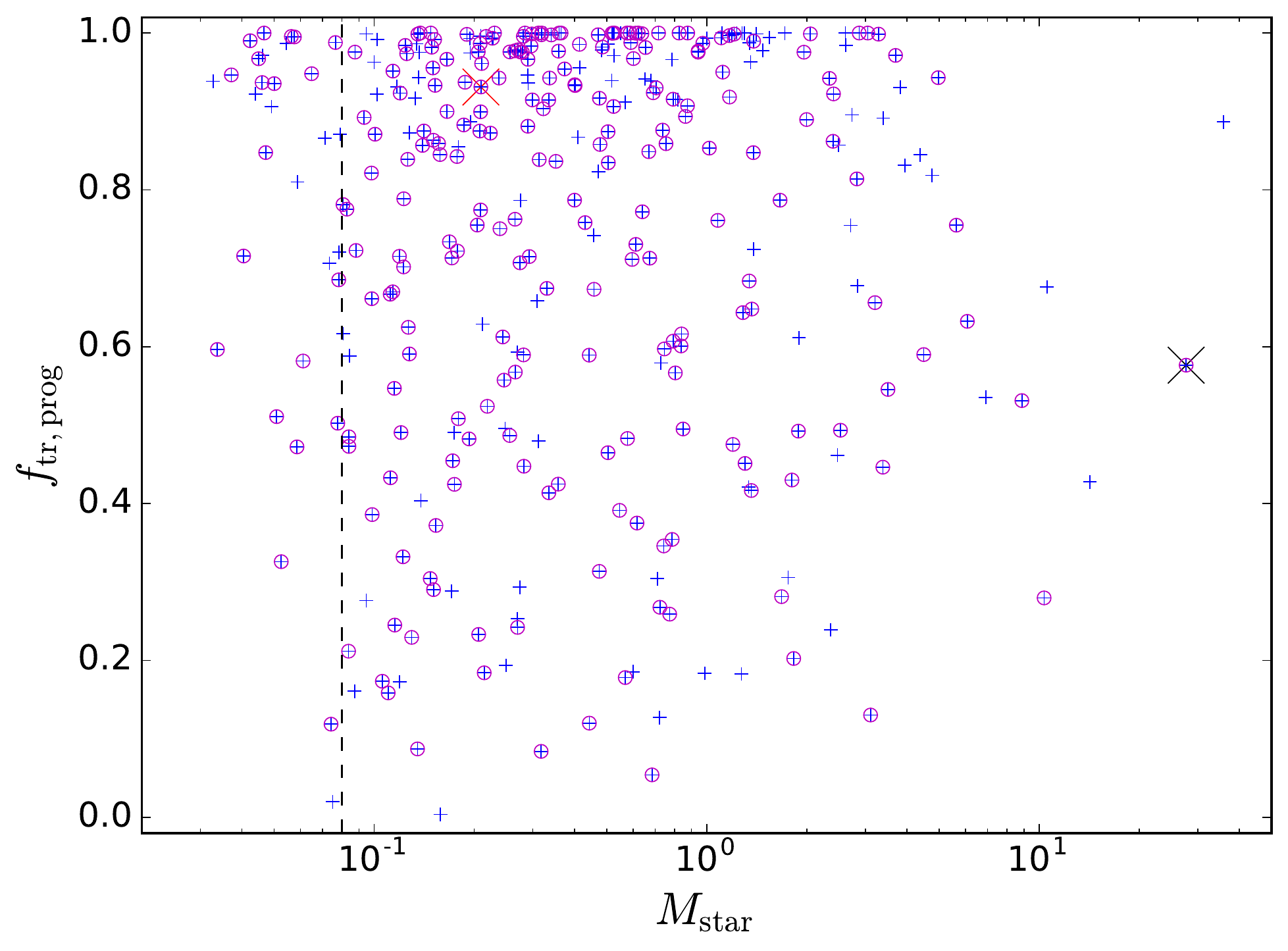}
   \includegraphics[width=\hsize]{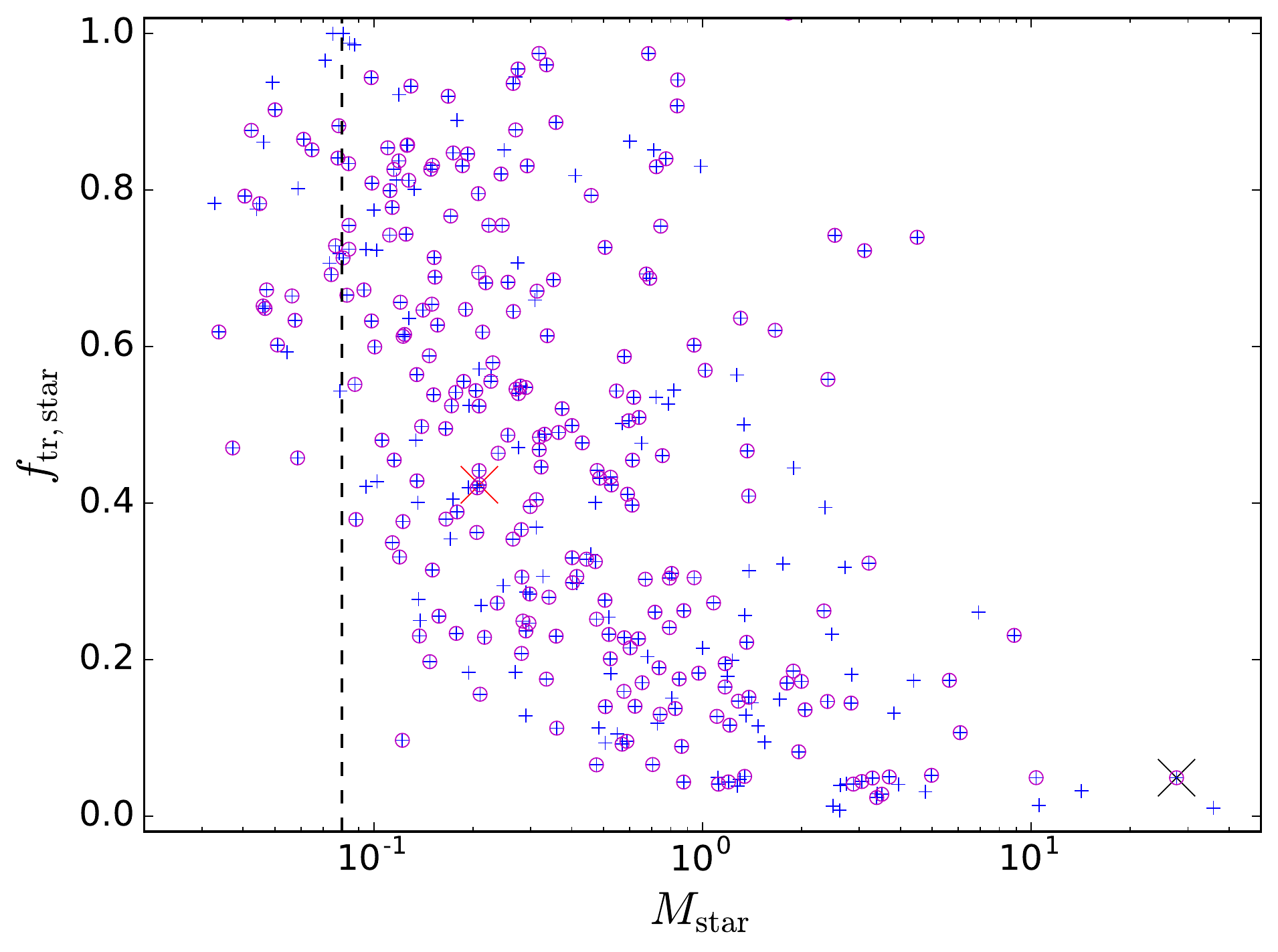}   
      \caption{Scatter plot of the progenitor mass fraction, $f_{\rm tr,star}$ (upper panel), and stellar mass fraction, $f_{\rm tr,star}$ (lower panel), as a function of the final stellar mass, $M_{\rm star}$. The blue pluses are for all stars, while magenta open circles are for the stars that have effectively finished accreting. The black and red crosses shows the stellar mass fraction of the example progenitor cores in Figure~\ref{tracerexample}, high-mass and low-mass respectively. The black dashed vertical line shows the limit of brown dwarfs, $M_{\rm star}=0.08 \; M_\odot$. 
              }
         \label{tracer_starmass}
   \end{figure}

We have found 1) a poor correlation between progenitor and stellar masses, 2) a mean dependence of stellar mass on core mass with a slope significantly shallower than unity, 3) a stellar mass reservoir that extends well beyond the limits of the gravitationally-bound prestellar core. These results are inconsistent with the main hypothesis of the core-collapse model, which is $M_{\rm star}\approx\epsilon_{\rm prog} M_{\rm prog}$ (with $\epsilon_{\rm prog}\lesssim 0.5$ and approximately independent of mass) for the simulation. Because most of the stellar mass comes from a larger and unbound region, rather than from a gravitationally-bound reservoir, the fundamental assumption of the IMF models of \citet{Hennebelle+Chabrier08imf} and \citet{Hopkins12imf} and their further developments \citep[e.g.][]{Hennebelle+Chabrier09imf,Hopkins13} is incorrect, and an understanding of the IMF based on those models is incomplete at best.

A similar conclusion was reached in \citet{Padoan+20SN_massive}, based on a star-formation simulation of a 250~pc region of the interstellar medium, driven by supernovae. However, in that work only high-mass stars are resolved, so the core-collapse model, as well as the competitive-accretion model, are ruled out only for the origin of massive stars. Here, instead, we show for the first time that the lack of correlation between core and stellar masses applies to the whole IMF, although it may be more extreme for the most massive stars. To further confirm this, we show in Figure~\ref{tracer_starmass} a scatter plot of the stellar mass fraction, $f_{\rm tr,star}$, as a function of the final stellar mass, $M_{\rm star}$. The plot shows that the stellar mass fraction that comes from the progenitor has a rather uniform distribution between approximately 0.05 and 1, irrespective of $M_{\rm star}$. There is only a slight trend with mass. Even at the IMF peak, $\sim 0.2 \, M_\odot$, there are stars with values as low as $f_{\rm tr,star} = 0.1$. For low-mass stars ($M_{\rm star} < 2 \; M_\odot$) with $f_{\rm tr,prog} > 0.5$, we find that the median value of $f_{\rm tr,star}$ is 0.48 (172 stars). For intermediate-mass stars (2 -- 5 $M_\odot$), the median value of $f_{\rm tr,star}$ is only 0.14 (13 stars). Thus, the main assumption of the core collapse model is clearly violated for a majority of stars at all masses, not only for the most massive stars. 

\subsection{Competitive Accretion}

\begin{figure}
   \centering
   \includegraphics[width=\hsize]{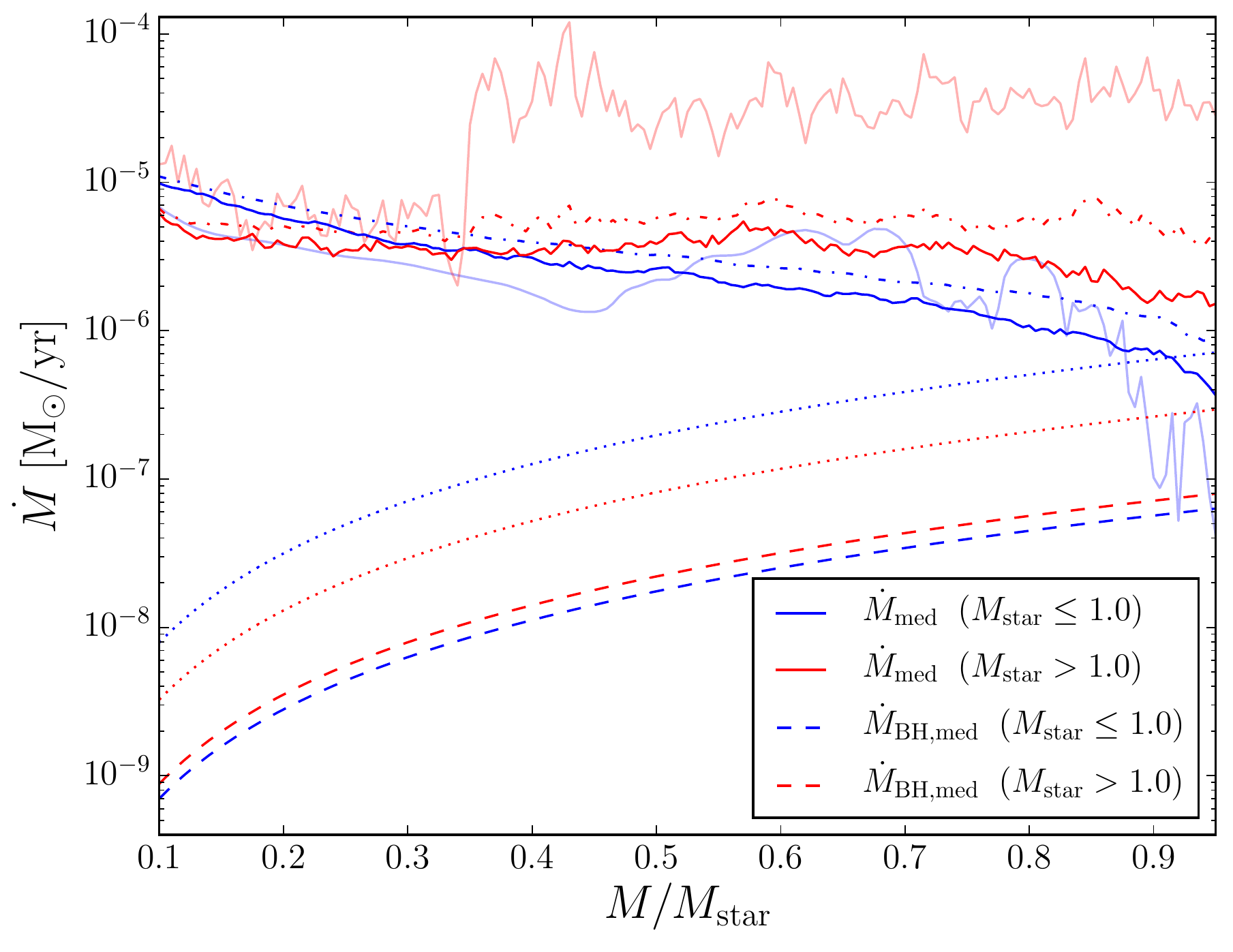}
      \caption{The time evolution of the accretion rate, $\dot{M}$, versus the time evolution of the stellar mass, $M$ (in units of the final stellar mass, $M_{\rm star}$), for stars that have finished accreting by the end of the simulation. The accretion rate evolution of individual stars have been stacked after interpolation onto the same values of final-mass fraction, to obtain a single average profile representing all the stars within two mass bins, $M_{\rm star} \leq 1 \; {\rm M_{\odot}}$ (blue lines) and $M_{\rm star} > 1 \; \rm M_{\odot}$ (red lines). Bright solid lines are median values (faint solid lines are for the two individual stars from Figure~\ref{tracerexample}), while dashed-dotted lines are mean values, dominated by the stars with the highest accretion rates. The predicted Bondi-Hoyle accretion rates, $\dot{M}_{\rm BH}$, are stacked in the same way, with dashed lines representing the median values, and dotted lines the mean values (see the text for the method of computing $\dot{M}_{\rm BH}$).  
              }
         \label{accretion_BH}
   \end{figure}

Although we find that a significant fraction of the stellar mass comes from outside of the progenitor cores, our simulation rules out also the competitive-accretion model \citep{Zinnecker82,Bonnell+2001a,Bonnell+2001b}. According to competitive accretion, the accretion rate of a star should increase with its mass as ${\dot M}\propto M^{2/3}$ in the case of gas-dominated potentials, or ${\dot M}\propto M^2$, in the case of stellar-dominated potentials \citep[e.g.][]{Bonnell+2001a,Bonnell+2001b}. Previous works that claimed to support competitive accretion, only showed scatter plots of accretion rate versus sink mass at a fixed time \citep{Maschberger+2014,Ballesteros-Paredes+2015,Kuznetsova+2018}, but no evidence that the accretion rate of a given star grows over time as the mass of that star increases. The time evolution of the accretion rates in our simulation is shown by the solid lines in Figure~\ref{accretion_BH}. After computing the time evolution of ${\dot M}$ and $M$ for every star, the profiles of ${\dot M}$ versus $M/M_{\rm star}$ (the stellar mass in units of its final value) are stacked together to obtain an average profile. This is done separately for stars with $M_{\rm star} \leq 1 \; {\rm M_{\odot}}$ (blue line) and $M_{\rm star} > 1 \; \rm M_{\odot}$ (red line). The accretion rate is relative constant as the stellar mass increases, for $M_{\rm star} > 1 \; \rm M_{\odot}$, or decreases with increasing mass for $M_{\rm star} \leq 1 \; {\rm M_{\odot}}$, in contrast with the prediction of competitive accretion. This confirms our earlier results in \citet{Padoan+14acc} and \citet{Padoan+20SN_massive}.

The average profiles in Figure~\ref{accretion_BH} appear to be relatively smooth, which is suggestive of analytical models of the collapse of an isothermal sphere \citep{Shu77,Dunham+14}. However, the profiles of individual stars (see the two examples shown by the faint blue and red lines in Figure~\ref{accretion_BH}) are much more irregular than the average ones, exhibiting fluctuations of one order of magnitude or larger that are inconsistent with the analytical collapse models. This is to be expected, because a significant fraction of the stellar mass originates well outside of the progenitor core. We have verified that only the mass that is initially found in the progenitor core is accreted in a way consistent with the collapse of an isothermal sphere, meaning with an accretion rate profile that is very smooth, relatively constant in time and approximately independent of the final stellar mass.

To further illustrate the departure from competitive accretion, we also estimate the Bondi-Hoyle accretion rate \citep{Bondi+44, Bonnell+2001a}, $\dot{M}_{\rm BH}$, using the current mass of the star, $M$, and the rms velocity, $v_{\rm 95}$, and the mean gas density, $\rho_{\rm 95}$, within the $R_{\rm 95}$ sphere of each sink particle at birth (the sphere with radius equal to $R_{\rm 95}$ and centered at the sink-particle position at birth): $\dot{M}_{\rm BH} = \epsilon_{\rm acc} 4 \pi G^2 M^2 \rho_{\rm 95} / (c_{\rm s}^2+v_{\rm 95}^2)^{3/2}$. This is just an order of magnitude estimate, as we account for the increase of the stellar mass over time, while the gas properties in the $R_{\rm 95}$ spheres are not varied with time. It is generally a conservative overestimate of the Bondi-Hoyle accretion rate, because $v_{\rm 95}$ is adopted as an estimate of the relative velocity between the star and the gas, without including the difference between the mean gas velocity in the $R_{\rm 95}$ sphere and the velocity of the star. The estimated rates are shown in Figure~\ref{accretion_BH}, using the same stacking procedure for the two groups of stars as before. The predicted Bondi-Hoyle accretion rates are on average a couple of orders of magnitude lower than the actual accretion rates of the sink particles in the simulation.  

\begin{figure}
   \centering
   \includegraphics[width=\hsize]{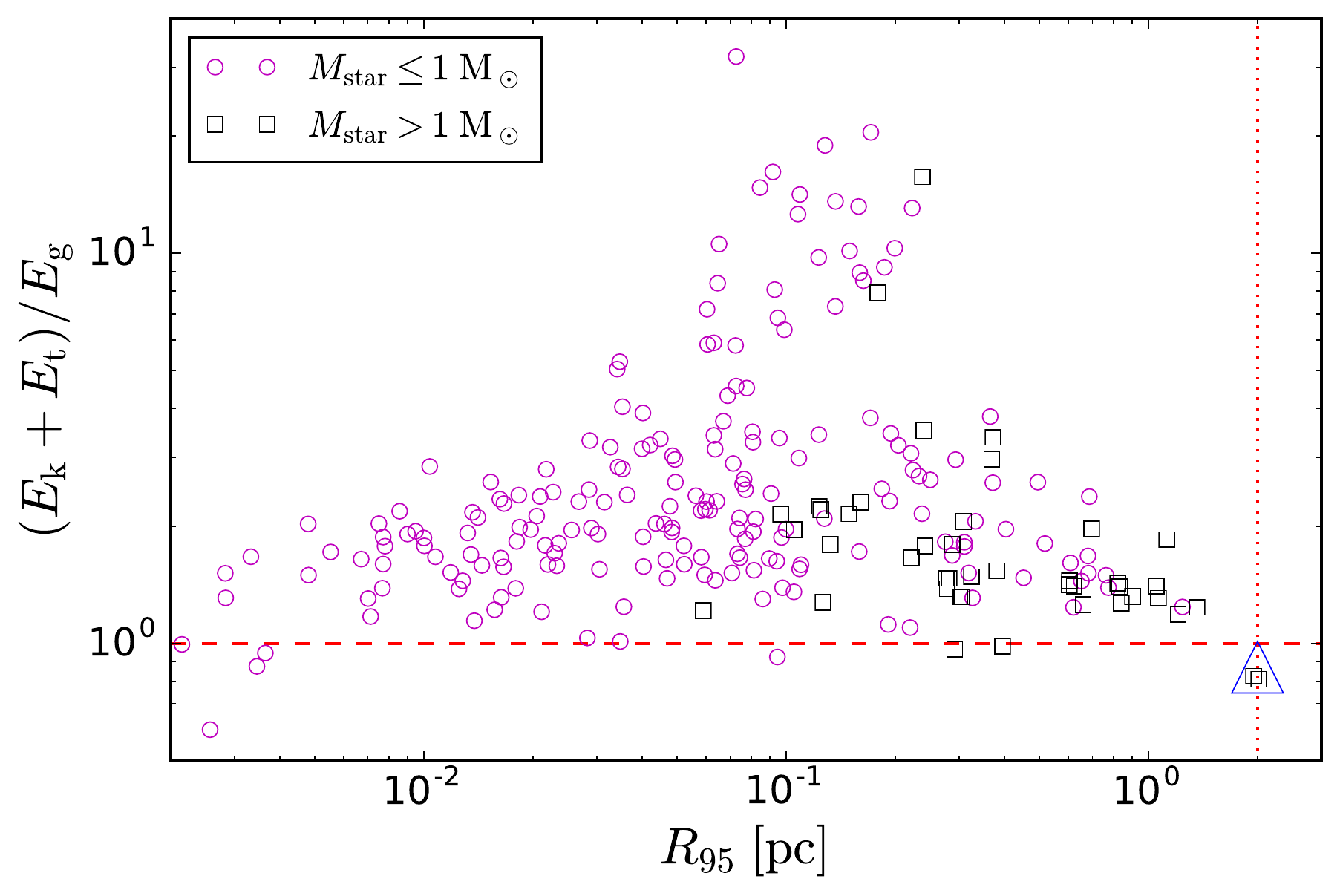}
      \caption{Scatter plot of the energy ratio, $E_{\rm k}+E_{\rm t}/E_{\rm g}$, of the spheres centered around the sink particles positions at birth and with radius equal to the inflow radius, $R_{\rm 95}$, versus $R_{\rm 95}$, shown for all stars that have finished accreting by the end of the simulation. Magenta circles and black squares are for $M_{\rm star} \leq 1 \; {\rm M_{\odot}}$ and $> 1 \; \rm M_{\odot}$, respectively. The blue triangle marks the energy ratio for the whole box, when the mass is calculated within a sphere of 2~pc radius rather than the whole simulation volume. The red dashed line marks the equipartition in the energy ratio, while the red dotted line marks the 2~pc size limit imposed by the 4~pc simulation cube.
              }
         \label{R95_boundness}
   \end{figure}

Low values of $\dot{M}_{\rm BH}$ and the failure of the competitive accretion model can be easily understood from the properties of the stellar mass reservoirs. The Bondi-Hoyle accretion rate is known to be too low if the virial parameter of the region containing the stellar mass reservoir is not much smaller than unity \citep[e.g.][]{Krumholz+05nature}. Figure~\ref{R95_boundness} shows that the sphere centered on the sink particle position at birth and with radius equal to the inflow radius, $R_{95}$, is almost always unbound. In the case of the two most massive stars, the values of $R_{95}$ are approximately half the size of the computational volume, so the energy ratio of their mass reservoirs is approximately the same as that of the whole simulation volume, shown by the triangle in Figure~\ref{R95_boundness}. Thus, the drop in the energy ratio with increasing $R_{95}$ for $R_{\rm 95}\gtrsim 1$~pc is a numerical constraint. With a larger computational volume, the outer scale of the turbulence would be much larger than 2~pc, and the mass reservoirs of the most massive stars could also be unbound. This is the case in the SN-driven simulation by \citet{Padoan+20SN_massive}, where the outer scale is $\sim 70$~pc.

\subsection{The Observed Core Mass Function}  \label{sect_obs}

We have identified the progenitor cores from the 3D density and velocity data-cubes of the simulation, and with the knowledge of when and where the star particles are formed. Although this allows us to test the main hypothesis of the core-collapse model, it does not closely resemble the process of identification of prestellar cores in the observations. The observed prestellar CMFs are extracted from 2D intensity information, such as dust-emission or dust-extinction maps, which involves a degree of line-of-sight confusion \citep[e.g.][]{Juvela+19}. Furthermore, the observed quantities must be converted into column density, which, in the case of sub-mm observations, depends on possible temperature and dust-opacity variations, leading to a significant uncertainty in the mass determination \citep[e.g.][]{Malinen+11,Roy+14,Pagani+15,Menshchikov16}. Resolution plays a role in deriving the CMF in our simulation, as shown in Appendix~\ref{app_parameters}: lowering the resolution by a factor of 8 results in an increase of the median mass by a factor of 3. A similar dependence of the CMF on resolution must also affect the observations. 

When a prestellar core is identified in the observations, it is hard to know how close it is to produce a protostar. The core may further increase in mass, or perhaps fragment into smaller cores, before the start of the collapse. In our study, we have selected the progenitors at the (approximate) time of star-particle formation. If we search for cores over the whole simulation box, using all available snapshots, and independent of the star-formation time (see Appendix~\ref{app_cmf}), we recover prestellar cores with median masses about a factor of 2 -- 3 higher than in the case of the progenitor cores at the time of star formation (Figure~\ref{comparepre}). We interpret this result as showing that, well before their collapse, prestellar cores are more massive, and later fragment into smaller units before they start to collapse into protostars. This is confirmed by the fact that the mass factor decreases over time (Figure~\ref{comparepre}). Further evidence of subfragmentation of the more massive cores has already been shown in Section~\ref{statistical}. A similar 2 -- 3 mass factor due to the fragmentation of cores over time may also affect the observed CMF, in particular at the high-mass end.

It should also be stressed that CMFs are often derived from dust-continuum data without a knowledge of the internal velocity dispersion in the cores, so the internal kinetic energy is neglected in the selection of gravitationally-bound cores. Beside the neglect of the kinetic energy, the core thermal and gravitational energies derived from the observations have significant uncertainties. To account for such uncertainties, observers often assume that a core is gravitationally bound, and can be classified as prestellar, if its estimated mass is half of the critical Bonnor-Ebert mass corresponding to the core radius. Besides the uncertainties in the mass determination mentioned above, the peak of the derived CMF is sensitive to this definition of prestellar cores, and would shift to larger masses if only gravitationally unstable cores (more massive than the critical Bonnor-Ebert mass) were selected, and/or the core internal kinetic energy were included. 

The issues we have raised with regards to the observed CMFs may explain the variation of the peak of the CMF from region to region, 
even within a homogeneous set of CMFs (same telescope, same data analysis, same research group). For example, the peak mass is found to be a few times larger than the IMF peak in Aquila and Orion \citep{Konyves+15,Konyves+2020_orion}, while it is very close to, or even slightly smaller than the IMF peak in Taurus and Ophiucus \citep{Marsh+16,Ladjelate+2020_ophiucus}. Despite the observational uncertainties and the significant differences in the way cores are extracted from the simulation and the observations, we do find a statistical similarity between the CMF and the IMF, as in the observations. However, our results show that, rather than looking at the IMF as being born from the CMF through a constant efficiency factor, both the CMF and the IMF should be viewed as being formed and fed by the same inertial inflows that arise naturally in supersonic turbulence. 

\section{Conclusions}

We have addressed the relation between the final mass of a star and that of its prestellar core to test the main hypothesis of the core-collapse model. Using a simulation that produces a realistic stellar IMF under realistic physical conditions found in molecular clouds, and hence goes well beyond the earlier investigations of \citet{Smith+09} and \citet{Gong+Ostriker15}, we have extracted, for each star particle, its gravitationally-bound progenitor core, at the time when the star is created. From a statistical analysis, a one-to-one comparison between progenitor and stellar masses, and a study of the mass flow based on tracer particles, we have reached the conclusions listed in the following.

\begin{enumerate}

    \item The progenitor CMF converges with resolution, with a peak moving from 0.66 M$_{\odot}$ to 0.28 M$_{\odot}$ using resolutions from 800~AU to 100~AU. The estimated converged position of the peak is 0.26 M$_{\odot}$, close to the IMF peak, which is contradictory to the core-collapse model.

    \item The CMF derived from the simulation is very similar to the stellar IMF from the same simulation. Irrespective of this statistical similarity, we find no direct correlation between the progenitor core mass and the final stellar mass for individual stars, contrary to the hypothesis of the core-collapse model.
    
    \item A significant fraction of the mass reservoir of stars is generally outside of the progenitor cores. This applies across the whole IMF, not just for massive stars. For stars less than 1 M$_\odot$, $\approx$50\% of the stellar mass originates outside the core. This increases to $\approx$90\% for intermediate-mass stars ($2 < M / M_\odot < 5$).
    
    \item The inflow region that contains 95\% of the mass reservoir of a star is generally much larger than the size of the progenitor core. The ratio between the inflow radius and the core radius has a median value of 14 and its largest values are $\sim10^3$. This size ratio shows no significant correlation with the final stellar mass. 
    
    \item The competitive-accretion model is also ruled out: the region that amounts to the stellar mass reservoir is not gravitationally bound, hence the Bondi-Hoyle accretion rate from that region is far too small to explain the actual accretion rate. On average, the accretion rate of a star is a decreasing function of its mass, contrary to competitive accretion.

    \item Observed CMFs should in principle result in a larger core mass on average, compared to the cores selected in the numerical data, because of resolution effects and the possibility that observed cores fragment prior to the star formation. Inclusion of cores that may not be gravitationally unstable in the observed CMFs may have the opposite effect.

\end{enumerate}

The main conclusion of this work is that the results from one of the highest dynamic range star-formation simulations to date, which resolves both the IMF and the CMF, rule out the core-collapse idea for all stellar masses. Because this idea is a fundamental assumption in some recent theoretical models of the stellar IMF, our work implies that attempts at understanding the stellar IMF based on this assumption are incomplete. Specifically, the similarities of the observed CMF with the stellar IMF should not be interpreted as a one-to-one relation between individual core masses and final stellar masses, as the two masses are poorly correlated even when prestellar cores are identified without all the uncertainties affecting the observations. Having now established that the stellar mass reservoir resides well beyond the limits of gravitationally-bound prestellar cores, future theoretical and observational work should address the role of inertial inflows in shaping the stellar IMF.

The main aspects of the results we present here are conceptual in nature, and are thus not likely to crucially depend on numerical resolution or simplifying assumptions such as using an isothermal equation of state. That said, access to even larger computational resources and more efficient computational methods will make it possible to follow up on this investigation by removing some of the computational limitations and omissions of physical effects. Candidates for such improvements include non-ideal MHD, and realistic cooling and heating, with diffuse and point-source radiative energy transfer, to name the perhaps most import ones.  In addition, the effective resolution that may be achieved with given resources also depends crucially on optimal use of adaptive mesh refinement, where a mix of criteria may well results in more optimal distributions of cells than relying entirely on the traditional Jeans's length criterion.

\section*{Acknowledgements}
We are grateful to the anonymous referee for a detailed and useful report.
PP and VMP acknowledge support by the Spanish MINECO under project AYA2017-88754-P, and financial support from the State Agency for Research of the Spanish Ministry of Science and Innovation through the "Unit of Excellence Mar\'ia de Maeztu 2020-2023" award to the Institute of Cosmos Sciences (CEX2019-000918-M).
The research leading to these results has received funding from the Independent Research Fund Denmark through grant No. DFF 8021-00350B (TH).
We acknowledge PRACE for awarding us access to Curie at GENCI@CEA, France.
Storage and computing resources at the University of Copenhagen HPC centre, funded in part by Villum Fonden (VKR023406), were used to carry out part of the data analysis. 

\section*{Data Availability}

Table 1, as well as other supplemental material, can also be obtained from a dedicated public URL (http://www.erda.dk/vgrid/core-mass-function/).

\bibliographystyle{mnras}
\bibliography{MC,padoan}

\begin{appendix}

\section{IMF convergence} \label{app_imf}

Our IMF results are in semi-quantitative agreement with those of \citet{Ballesteros-Paredes+2015} and \citet{Federrath+2017}, while the much larger mass scale obtained by \citet{Guszejnov+2020} is incompatible with our results as well as with observations. Using a cloud with 1000 solar masses of gas concentrated into an isolated Bonnor-Ebert like structure with radius 0.084 pc, \citet{Lee+2018b} find convergence of the IMF only after adding non-isothermal effects. Due to the absence of larger scales, which precludes the existence of a quasi-stationary state, these results are not directly comparable to the much more extended (in space and time) simulations referenced above. Earlier results by e.g.\ \citet{Girichidis+2011}, and \citet{Krumholz+12imf} also generally span too small regions in space-time to be directly comparable with more recent simulations.
   
\begin{figure}
   \centering
   \includegraphics[width=\hsize]{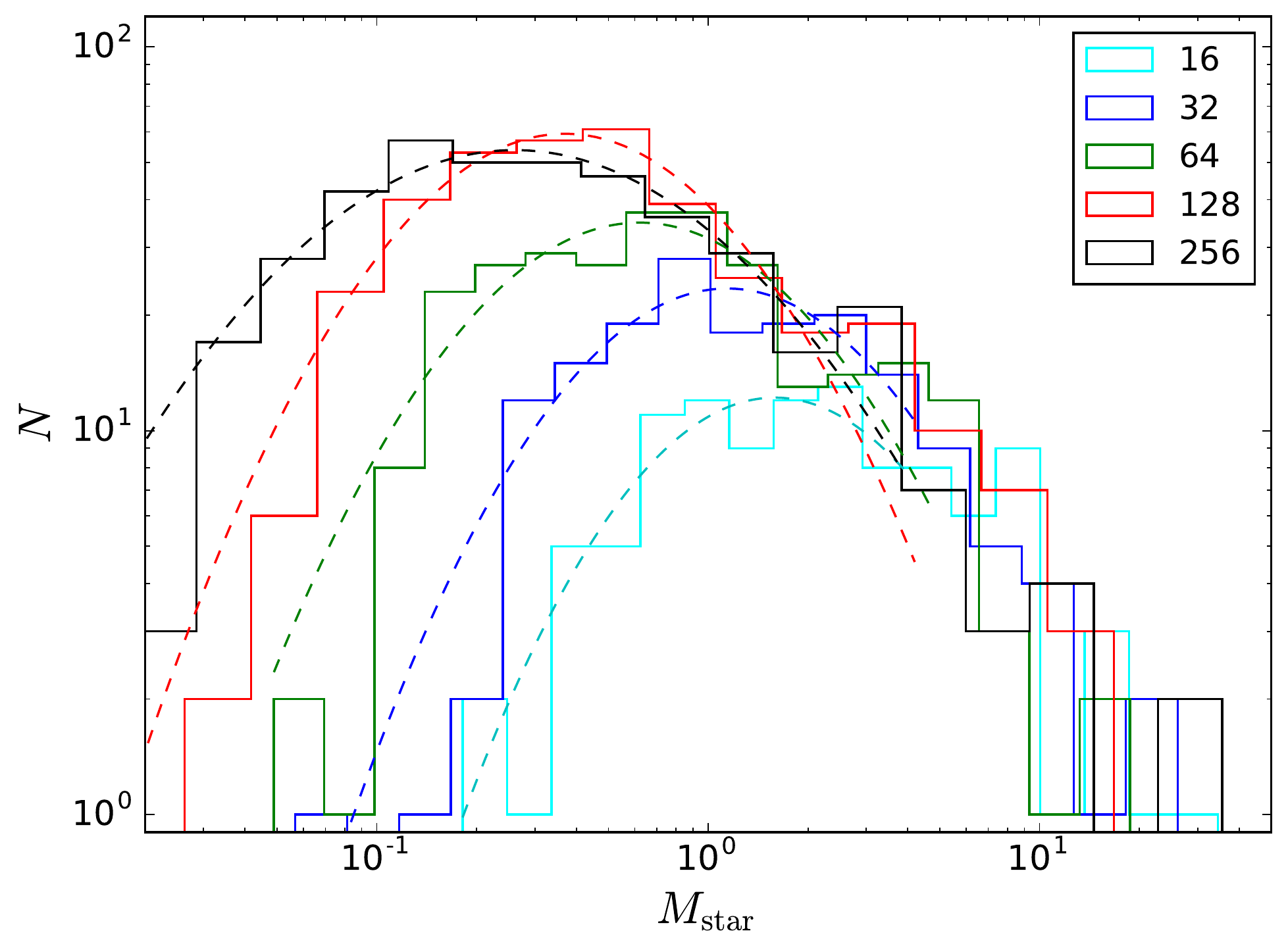}
   \includegraphics[width=\hsize]{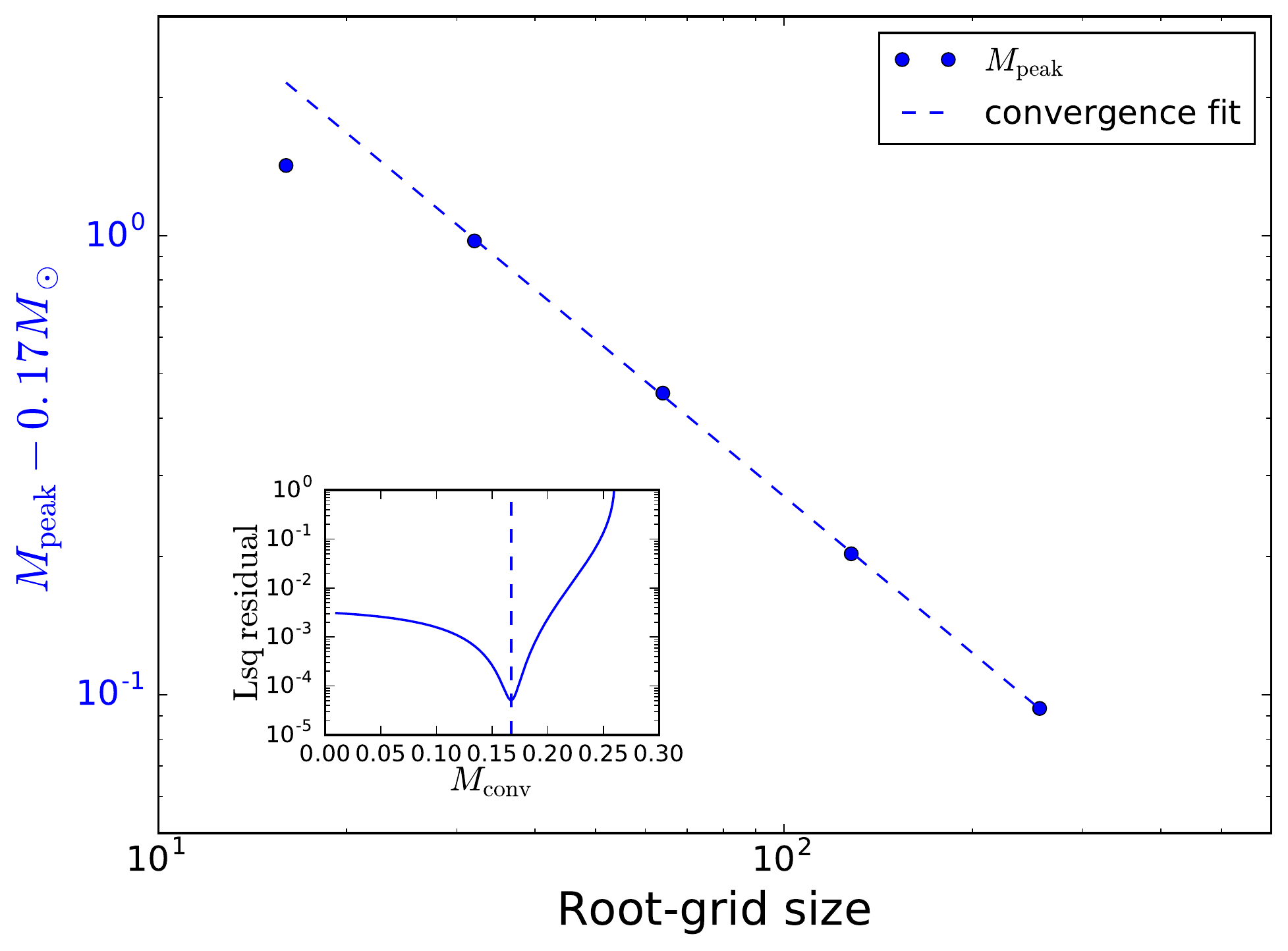}
   
      \caption{{\it Top:} Stellar IMF histograms for different MHD runs with root-grid sizes $16^3$, $32^3$, $64^3$, $128^3$ and $256^3$. Dashed lines are lognormal fits ($N \propto \exp(-(\log_{10} M - \log_{10} M_{\rm peak})^2 / 2 \sigma^2$) to masses lower than $4 \; M_\odot$, and the values are given in Table~\ref{imftable}. {\it Bottom:} The peak of the IMF as a function of the simulation root-grid size. The peak of the IMF converges towards $M_{\rm conv}=0.17 \; M_\odot$, as shown by $M_{\rm peak}(x)-M_{\rm conv}$ almost forming a straight line (blue dashed line) in logarithmic space. The $16^3$ resolution is not included in the convergence fitting. The inset shows the least-squares fitting residuals, with the minimum marked by a vertical dashed blue line.
              }
         \label{compareimf}
   \end{figure}

The simulations of \citet{Haugbolle+18imf} showed that the IMF peak in isothermal MHD turbulence tends to a converged value. Here, we demonstrate again the numerical convergence of the IMF peak in the simulations of \citet{Haugbolle+18imf} using a different method that allows us to estimate the converged value. The results are shown in Figure~\ref{compareimf}. The top panel shows the stellar IMF histograms from simulations with different root grid sizes \citep[for details, see][]{Haugbolle+18imf}. The bottom panel shows the peak of the IMF minus the converged value as a function of the simulation root-grid size. A convergence of the peak of the IMF is shown by these points forming almost a straight line (dashed blue line) in logarithmic space. The converged value of the peak of the IMF is found by minimizing the least-squares fit residuals of the points from the straight-line fit, by iterating through a mass range. This results in the converged IMF peak value of $0.17 \; M_\odot$, shown in the inset. The lowest root grid size, $16^3$, is not included in the convergence fitting, as its peak value is not fitted well even with a mass limit of $4 \; M_\odot$, as can be seen from its histogram in the top panel. Lowering the mass limit to $2 \; M_\odot$ requires dropping the $32^3$ case from the fitting, but the remaining three higher root grid sizes points fit almost the same converged peak value as before.  

\begin{table}
\caption{Parameters of the lognormal fits to mass histograms in the top panel of Figure~\ref{compareimf}, and the number of stars, $N_{\rm star}$, in each histogram. }           
\label{imftable}      
\centering                          
\begin{tabular}{c c c c c}        
\hline\hline                 
Resolution & $M_{\rm peak}$ [$M_\odot$] & $\sigma$ & $N_{\rm star}$ \\ 
\hline                        
16 &  1.59 &  0.42 & 107 \\
32 &  1.14 &  0.45 & 170 \\
64 &  0.62 &  0.48 & 278 \\
128 &  0.37 &  0.47 & 364 \\
256 &  0.26 &  0.60 & 411 \\
\hline                                   
\end{tabular}
\end{table}

\section{Surface terms}\label{app_surface}

Following the formulation of the Eulerian Virial Theorem \citep[][]{Parker79, McKee+Zweibel92, McKee99} by \citet{Dib+07}, the kinetic, $S_{\rm k}$, and thermal, $S_{\rm t}$, surface terms are defined as:
\begin{equation}
S_{\rm k} = \frac{1}{2} \oint_S r_i \rho v_i v_j \hat{n_j} dS,\\
S_{\rm t} = \frac{3}{2} \oint_S r_i P \hat{n_i} dS,
\label{eqsurface}
\end{equation}
where $\mathbf{r}$ is the distance from the center-of-mass, $\rho$ is density, $\mathbf{v}$ is the velocity relative to the center-of-mass, $\mathbf{\hat{n}}$ a normal unit vector on surface element $dS$, and $P$ is thermal pressure.

Given the highly complicated shape of some of our cores, we approximate the surface terms in the following manner. We go over each cell in the progenitor core and flag it as a border cell if at least one of its six face-on neighbors is not part of the core. We then calculate the surface terms by summing over the border cells:
\begin{equation}
S_{\rm k} = \frac{A_{\rm cell}}{2} \sum r_n \rho_n v_{r,n}^2,\\
S_{\rm t} = \frac{3 A_{\rm cell}}{2} \sum r_n P_n,
\label{eqskst}
\end{equation}
where $A_{\rm cell}$ is the area of one face of a cell, $r_n$ is the distance of cell n from the sink particle, $\rho_n$ is the density of cell n, $v_{r,n}$ is the radial component of the velocity of cell n after the subtraction of the mean velocity of the core, and $P_n$ is the thermal pressure of cell n.

In addition, we also calculate the kinetic energy of the collapse motion towards the sink particle, $E_{\rm coll}$:
\begin{equation}
S_{\rm k} = \frac{V_{\rm cell}}{2} \sum \rho_n v_{r,n}^2,
\end{equation}
where $V_{\rm cell}$ is the volume of a single cell.

\begin{figure}
   \centering
   \includegraphics[width=\hsize]{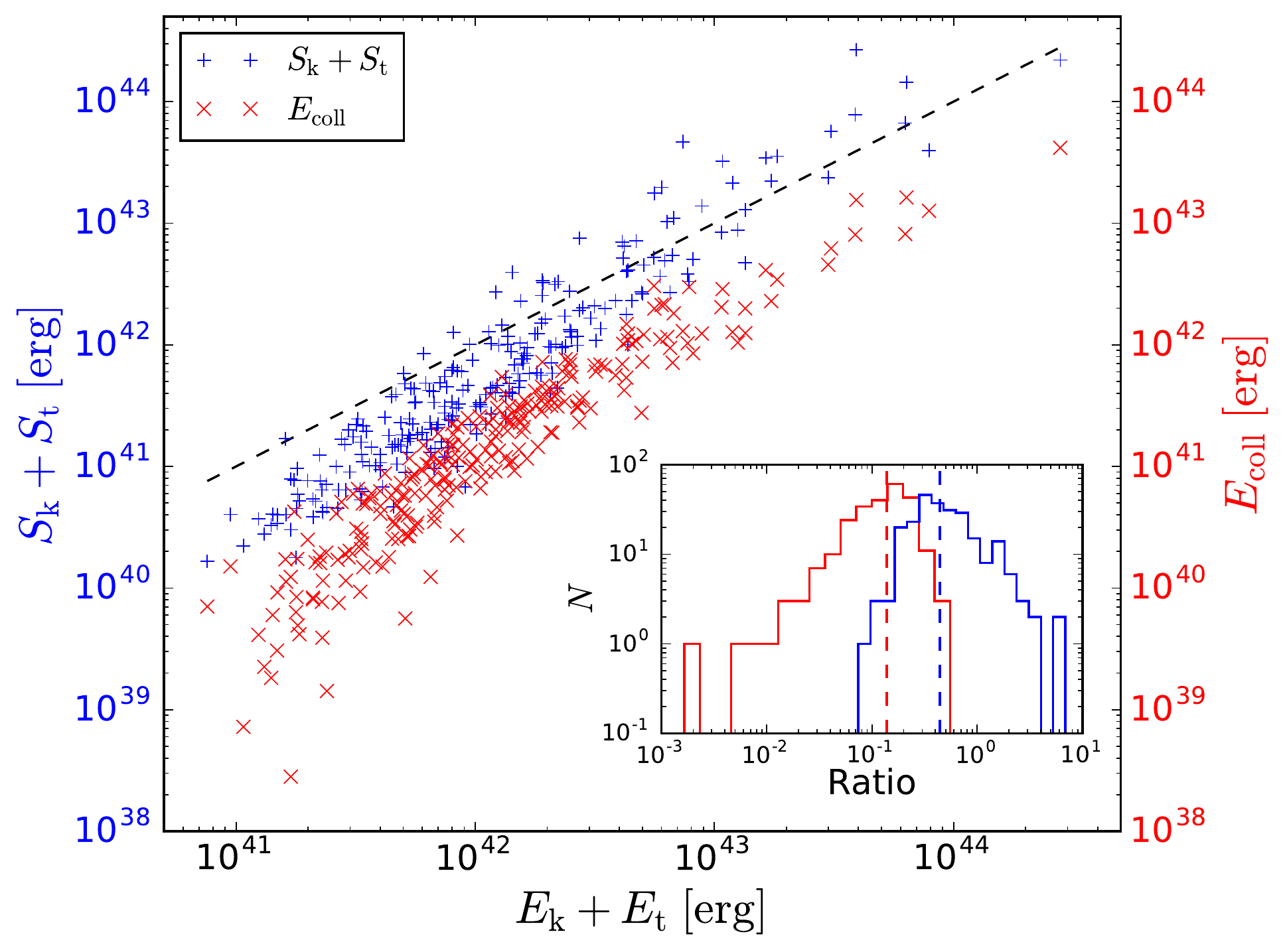}
      \caption{Scatter plot of the sum of surface terms, $S_{\rm k} + S_{\rm t}$ (blue plusses), and the kinetic energy of the collapse motion, $E_{\rm coll}$ (red crosses), against the sum of the internal energy, $E_{\rm k} + E_{\rm t}$. Black dashed line signals 1-to-1 relation. The inset shows the histogram of the ratio of the surface terms to the internal energy (blue line) and the ratio of $E_{\rm coll}$ to the internal energy (red line). Dashed lines are at the median values, 0.44 and 0.14, respectively.
              }
         \label{surface}
   \end{figure}

Figure~\ref{surface} shows the surface terms and the kinetic energy of the collapse motion against the sum of internal kinetic and thermal energies, $E_{\rm k}$ and $E_{\rm t}$. The inset shows the histograms of the ratios, showing that the median values of the contribution of the surface terms is 44\% and of the collapse motion 14\%.

\section{Clumpfind parameters}\label{app_parameters}

Although the use of density isosurfaces and the value of the related parameters in the clumpfind algorithm are somewhat arbitrary, the tests reported here demonstrate that the main results are insensitive to the details of the choice of parameters.

The clumpfind lower limit of 4 cells per core has no significant effect on the results, since while the selection is done at 100 AU resolution, the MHD run goes down an additional step in resolution. The 4 clumpfind cells thus correspond to 32 MHD cells. Only 12 cores (3.5\%) have less than 100 such cells, and excluding those cores would not have a noticeable effect on the results. Only 62 cores (18\%) have less than 1000 cells, and excluding even those cores would not have a significant effect.

\begin{figure}
   \centering
   \includegraphics[width=\hsize]{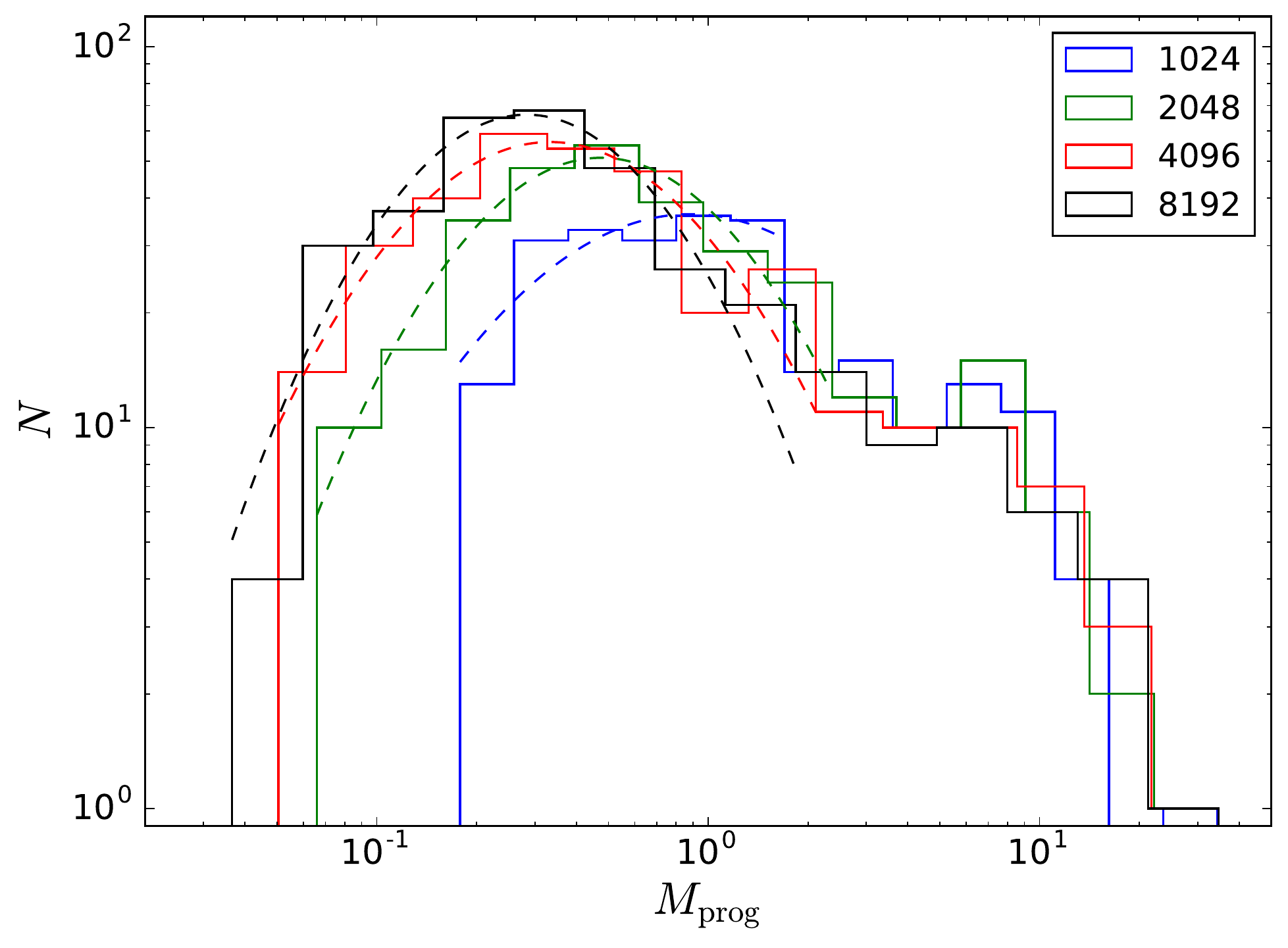}
   \includegraphics[width=\hsize]{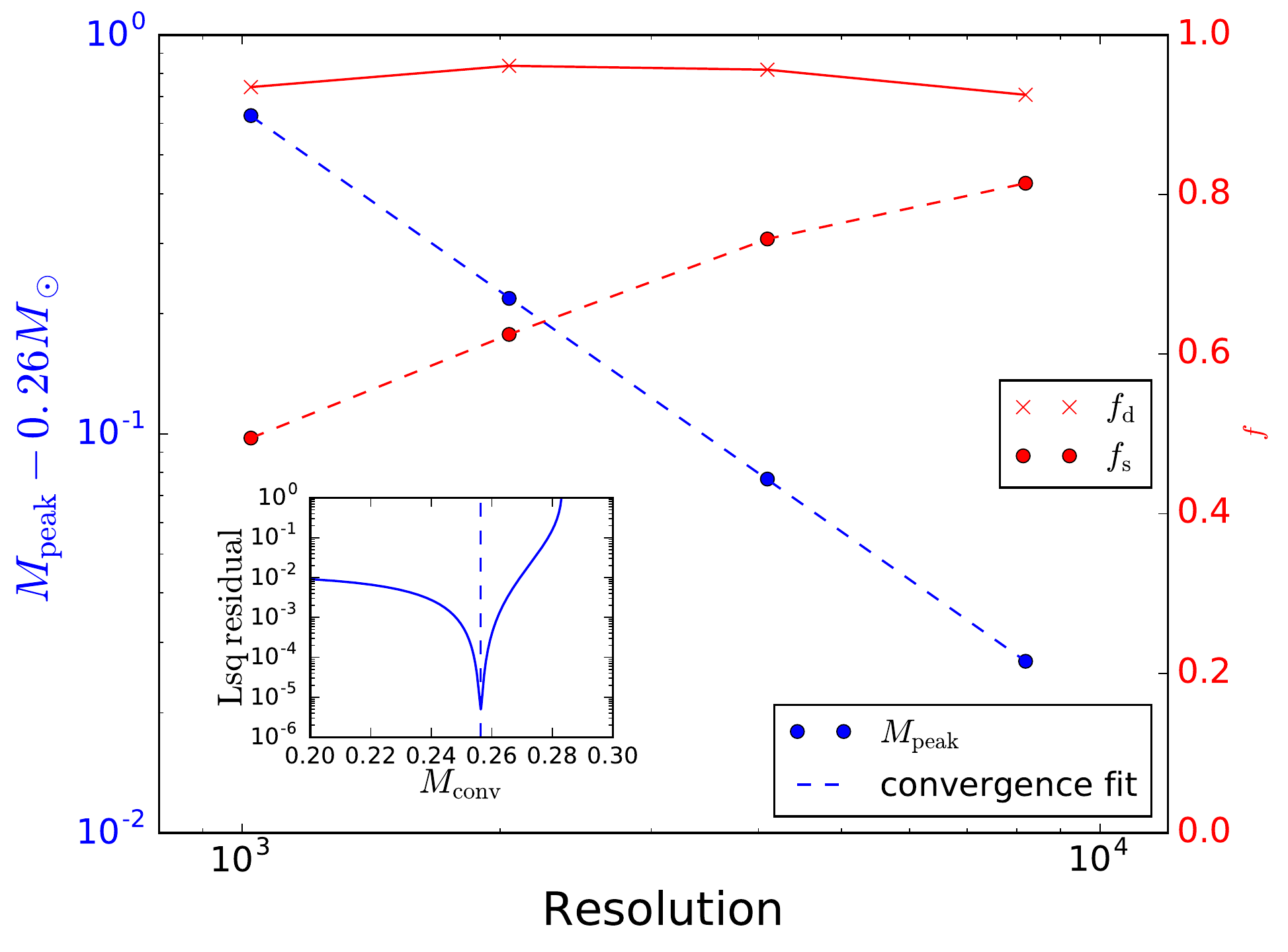}

      \caption{{\it Top:} Progenitor CMFs for resolutions $1024^3$, $2048^3$, $4096^3$ and $8192^3$ for the whole box, using $f=2\%$ and 0.5~pc subcubes around the star particles in the selection. Dashed lines are lognormal fits ($N \propto \exp(-(\log_{10} M - \log_{10} M_{\rm peak})^2 / 2 \sigma^2$) to masses lower than $2 \; M_\odot$, and the values are given in Table~\ref{restable}. {\it Bottom:} The peak of the core mass function, the fraction of detected stars of all the 413 stars, and the fraction of single star progenitors of the detected stars, as a function of the full box resolution. The inset shows the least-squares fitting residuals, with the minimum marked by a vertical dashed blue line. The peak of the CMF converges towards $M_{\rm conv}=0.26 \; M_\odot$, as shown by $M_{\rm peak}(x)-M_{\rm conv}$ forming an almost straight line (blue dashed line) in logarithmic space.
              }
         \label{compareres}
   \end{figure}

The effect of the resolution is shown in Figure~\ref{compareres}. The top panel shows the CMF at different resolutions, using 0.5~pc subcubes around the star particles and $f=2\%$. It is clear that the CMF peak moves to lower masses with increasing resolution, as summarized in Table~\ref{restable}. The bottom panel shows how the peak of the progenitor CMF ($M_{\rm peak}$, blue line), the fraction of the detected stars ($f_{\rm d}$, red line) and the fraction of single-star progenitors ($f_{\rm s}$, dashed red line) depend on the resolution, as also reported in Table~\ref{restable}. The estimated value of the converged peak mass, $M_{\rm conv}=0.26 \; M_\odot$, is determined by the least-squares fitting minimum (see Appendix~\ref{app_imf}), when $M_{\rm peak}(x)-M_{\rm conv}$ is best approximated by a straight line in logarithmic space (dashed blue line), where $x$ is the resolution. The least-squares fitting residuals are shown in the inset, with the minimum residual value marked by a vertical dashed blue line. The bottom panel of Figure~\ref{compareres} also shows that $f_{\rm s}$ increases with resolution, which is another reason to use the highest feasible resolution. 

By contrast, the density ratio, $f$, does not seem to have an effect at the highest resolution, $8192^3$. The CMFs are almost identical for $f = 2\%, 4\%, 8\%$. This is presumably because, at $8192^3$, the core boundary is already sharp enough that we do not need to have a lower value of $f$ to separate neighboring cores from each other. 

\begin{figure}
   \centering
   \includegraphics[width=\hsize]{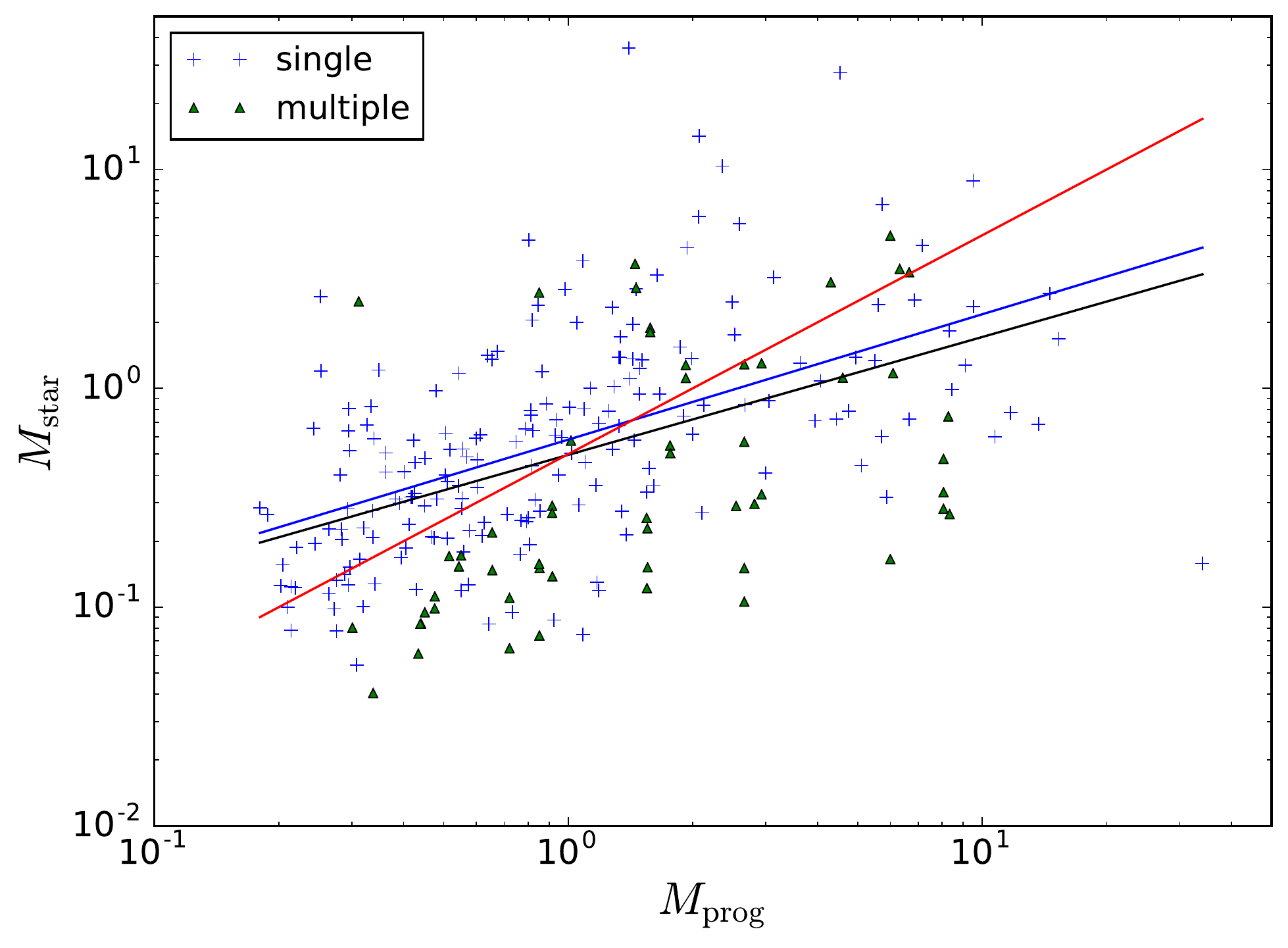}
      \caption{Same as the bottom panel in Fig.~\ref{star_to_progenitor_mass_fit}, but the core selection is done at $1024^3$ whole box resolution in 0.5~pc subcubes around the star particles. Progenitors are plotted with different symbols based on how many and how old stars they have inside them: progenitors with only one star (blue crosses), and progenitors that have multiple stars born at the same time (green triangle). The red line is drawn at $M_{\rm star}=\epsilon_{\rm acc} M_{\rm prog}$, with $\epsilon_{\rm acc}=0.5$. Black and blue lines are fits to all progenitors and the single-star progenitors, respectively. 
              }
         \label{1024_multiplicity}
   \end{figure}

Figure~\ref{1024_multiplicity} shows the one-to-one scatter plot of final stellar masses and the masses of their progenitors, like in the bottom panel of Fig.~\ref{star_to_progenitor_mass_fit}, but the selection is done at $1024^3$ resolution instead of $8192^3$. It is clear that the one-to-one correlation of progenitor and stellar masses is even worse than in Fig.~\ref{star_to_progenitor_mass_fit}, giving a value of the Pearson's correlation coefficient $r = 0.48$, even though the slope is similar, $a=0.54$. However, considering only the progenitors with a single newborn star, we get $r = 0.53$ and $a=0.57$ , which are similar to those found for single-star progenitors in Fig.~\ref{star_to_progenitor_mass_fit}, $r=0.55$ and $a=0.54$.
   
   \begin{table}
\caption{Parameters of the lognormal fits to mass histograms in the top panel of Figure~\ref{compareres}, and the fractions of detected progenitors, $f_{\rm d}$, and of single-star progenitors out of the detected ones, $f_{\rm s}$, used in the bottom panel of Figure~\ref{compareres}.  }           
\label{restable}      
\centering                          
\begin{tabular}{c c c c c}        
\hline\hline                 
Resolution & $M_{\rm peak}$ [$M_\odot$] & $\sigma$ & $f_{\rm d}$ & $f_{\rm s}$\\ 
\hline                        
1024 &  0.66 &  0.43 &  0.93 &  0.49 \\
2048 &  0.46 &  0.37 &  0.96 &  0.62 \\
4096 &  0.34 &  0.43 &  0.96 &  0.74 \\
8192 &  0.28 &  0.42 &  0.92 &  0.81 \\
\hline                                   
\end{tabular}
\end{table}
   
\section{The CMF of the whole box} \label{app_cmf}

What does the CMF look like, if we do the detection on the whole 4~pc simulation box without {\it a priori} knowledge of the star particles? We add the stellar mass in (see Sect.~\ref{detection}) and do the bound core selection in each snapshot at $1024^3$ resolution, using minimum density level 10 and $f=2\%$. We then categorize them as prestellar cores without stars, prestellar cores with stars (one or more stars that formed in that snapshot) which would be our progenitor cores, and protostellar cores (one or more older stars), according to the star locations and ages. 

\begin{figure}
   \centering
   \includegraphics[width=\hsize]{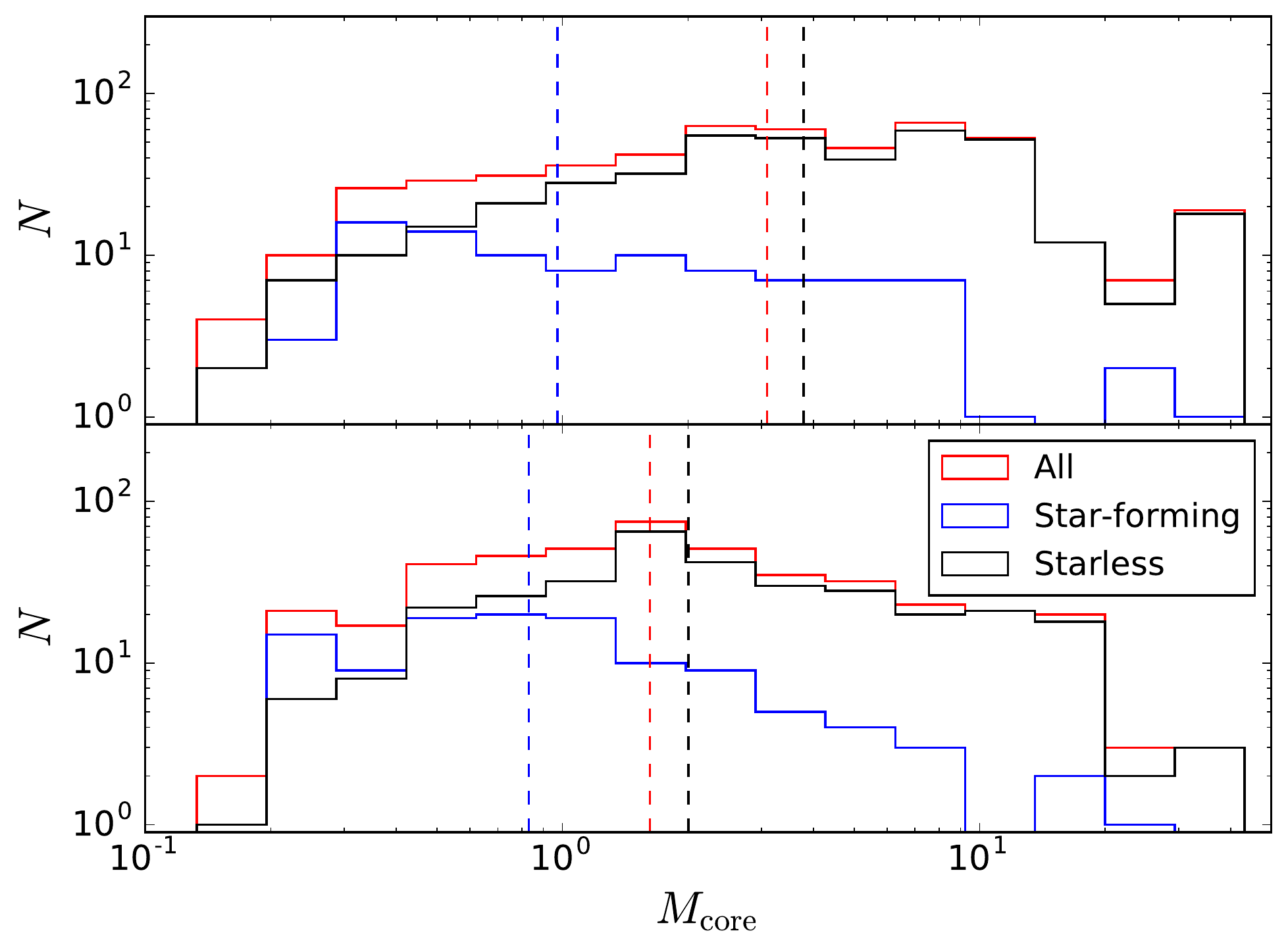}
      \caption{Mass histograms for all prestellar cores detected in the first half ({\it top panel}, 59 snapshots) and second half of the simulation ({\it bottom panel}, 59 snapshots), at resolution $1024^3$ and using $f=2\%$ in the detection. Detection is done for the whole 4~pc cube at once in each snapshot. We split all prestellar cores (red) between the progenitor cores where we know a star is forming (blue) and the ones currently not forming a star (black). The vertical dashed lines mark the median value for each selection. It can clearly be seen that while the median mass of the star-forming cores (blue dashed line) changes only a little, the median mass of the starless cores (black dashed line) decreases dramatically. The proportion of star-forming cores also grows from 19\% to 27\% of the all the cores detected. The ratio of median masses of all the cores (red dashed line) and the star-forming cores (blue dashed line) changes from 3 to 2.
              }
         \label{comparepre}
   \end{figure}

Figure~\ref{comparepre} shows the mass histograms of all prestellar cores detected in the 118 snapshots of the simulation, divided between the first half and the second half of the simulation, and in subcategories of prestellar cores without stars and progenitors where a star is currently forming. We can see from the median values that if we do not have {\it a priori} information on current star formation, the median mass of all prestellar cores is about three times higher than that of just the progenitors for the first half, but this decreases to a ratio of two in the second half, hinting that the larger bound clouds fragment into smaller ones as the simulation progresses. 

\section{Progenitor column density maps} \label{app_maps}

Column density maps for progenitors in Table~\ref{tableprog}, identified by the number in the upper left corner, with the letter based on the line-of-sight axis. Each map is based on the density subcube used in the identification of the progenitor, usually a 0.5~pc region (see the scale bar on the bottom left corner), centred on the newly-born star, with the time of the identification in the bottom right corner. The logarithmic colorbar is the same as in Figure 1, running from $N_{\rm H_2} = 10^{21.5} \; {\rm cm^{-3}}$ to $N_{\rm H_2} = 10^{24} \; {\rm cm^{-3}}$ before saturating. Cyan contours are drawn where there is at least one cell belonging to the progenitor core on the line of sight. Similar maps for all progenitors at even higher resolution are available at URL: http://www.erda.dk/vgrid/core-mass-function/.

\begin{figure*}
\centering
\includegraphics[width=\hsize]{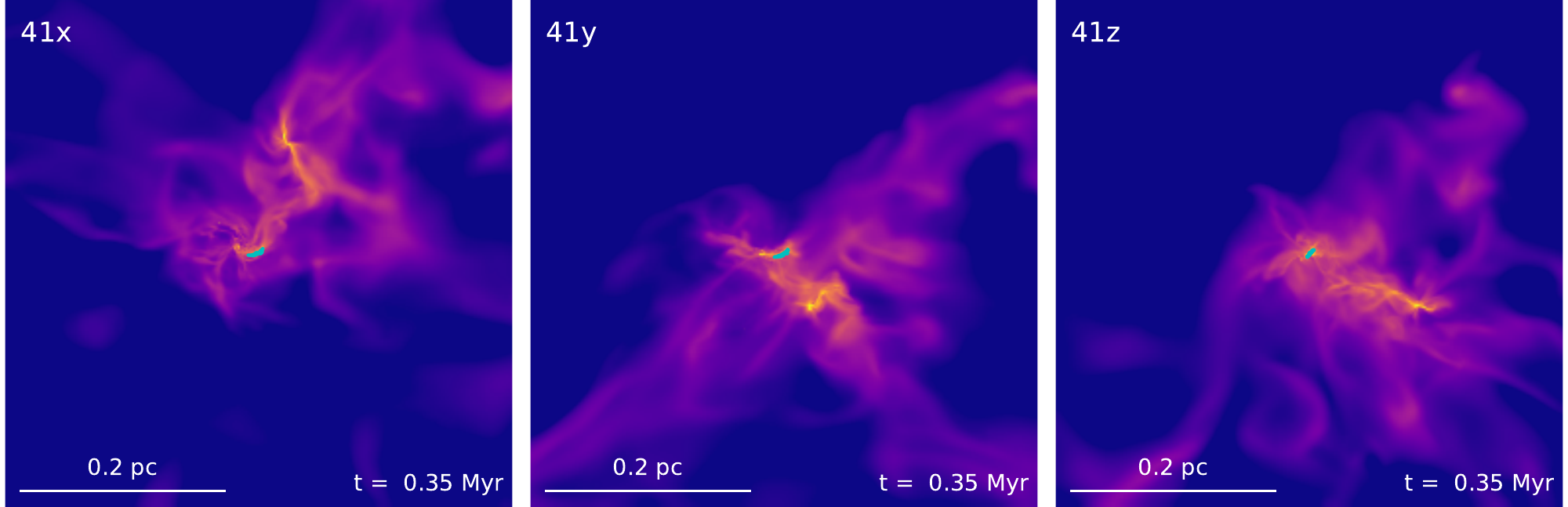}
\includegraphics[width=\hsize]{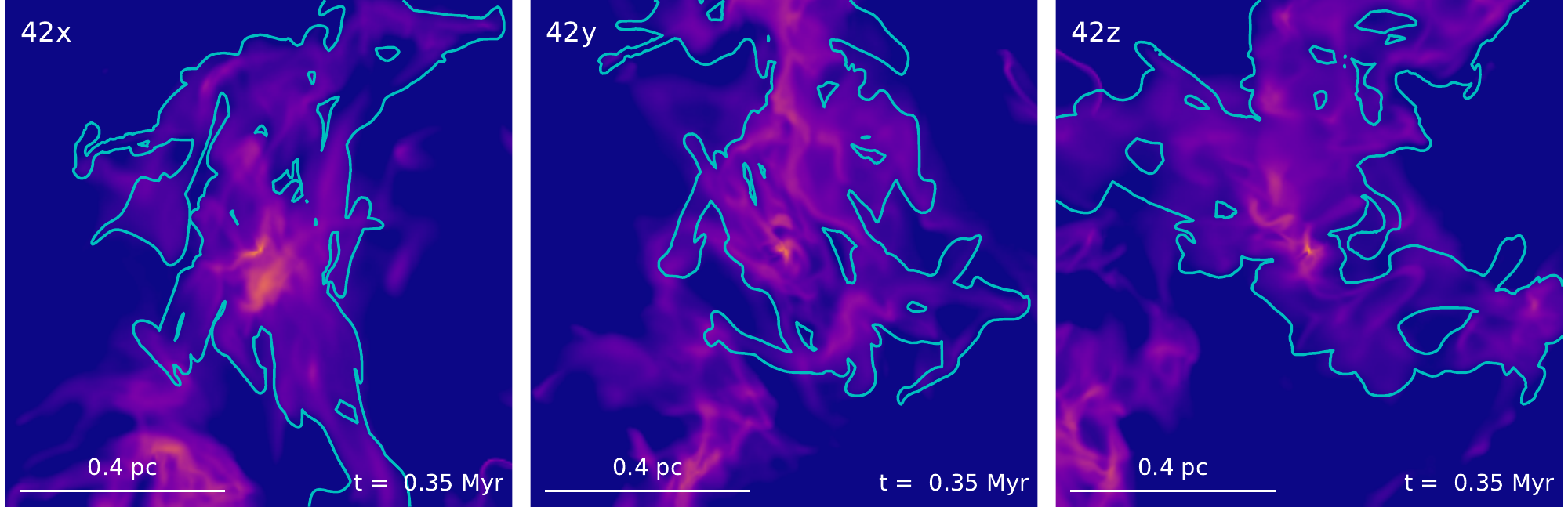}
\includegraphics[width=\hsize]{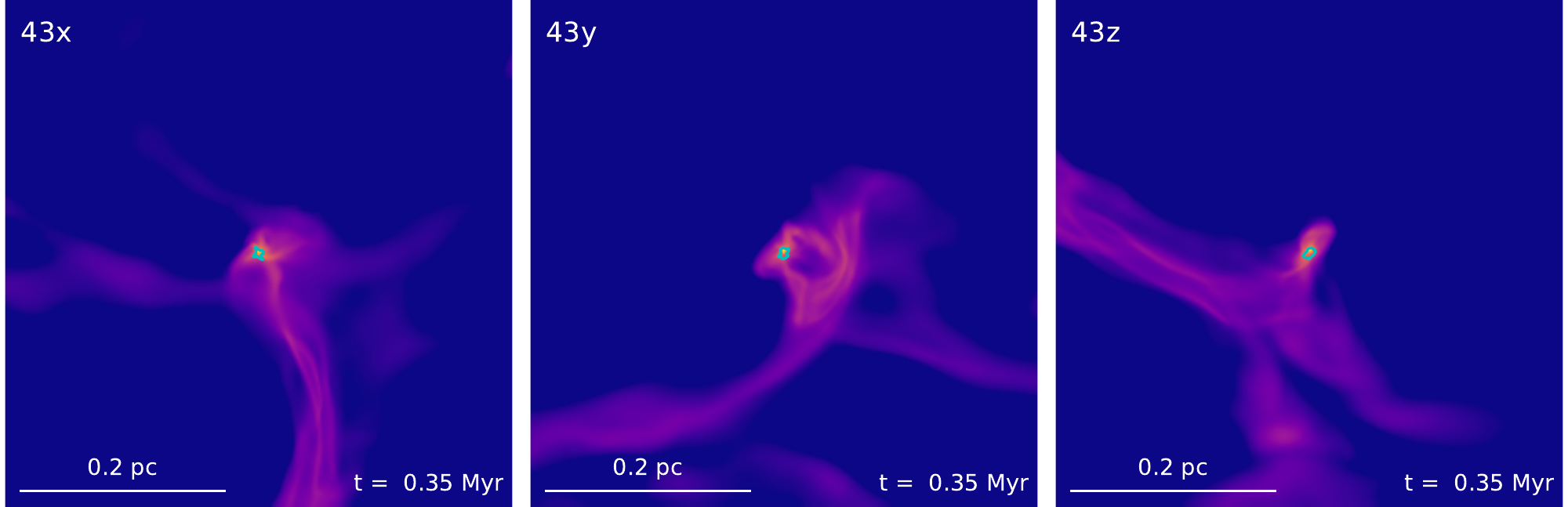}
\includegraphics[width=\hsize]{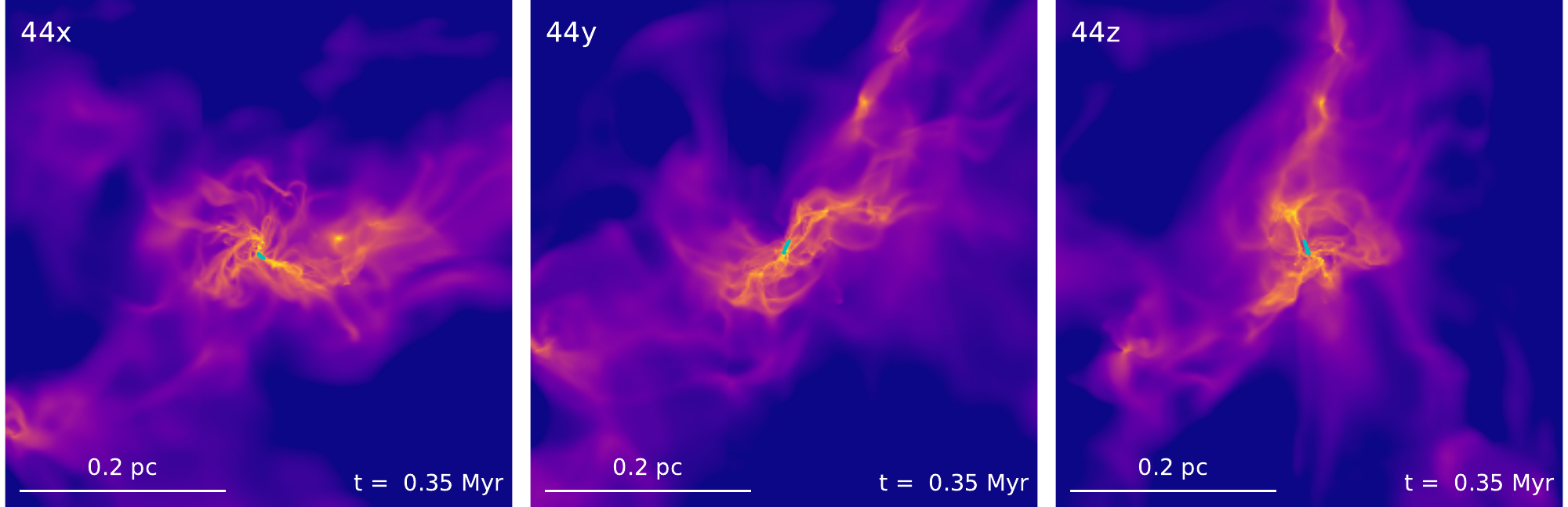}
\end{figure*}

\begin{figure*}
\centering
\includegraphics[width=\hsize]{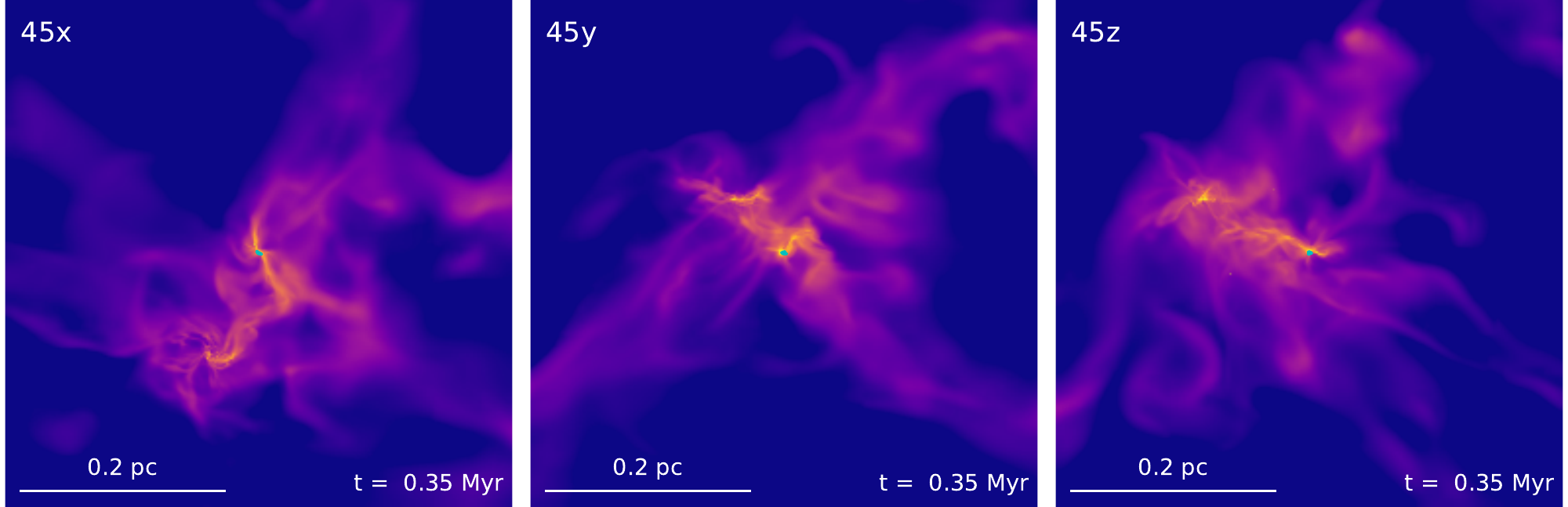}
\includegraphics[width=\hsize]{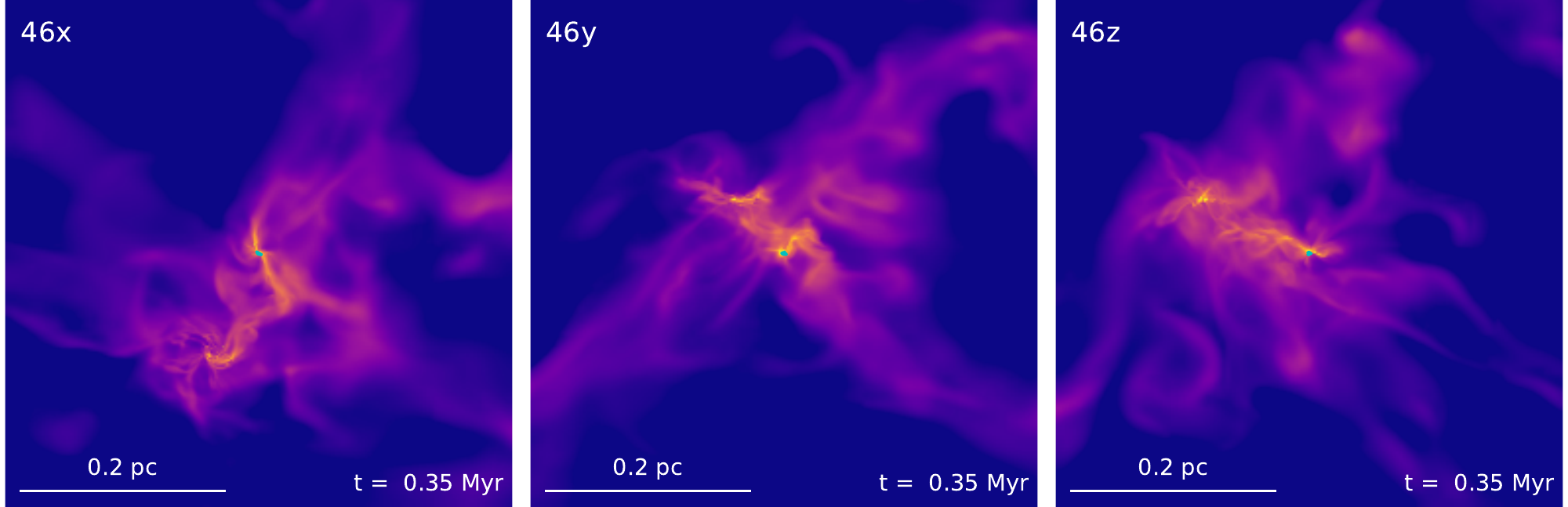}
\includegraphics[width=\hsize]{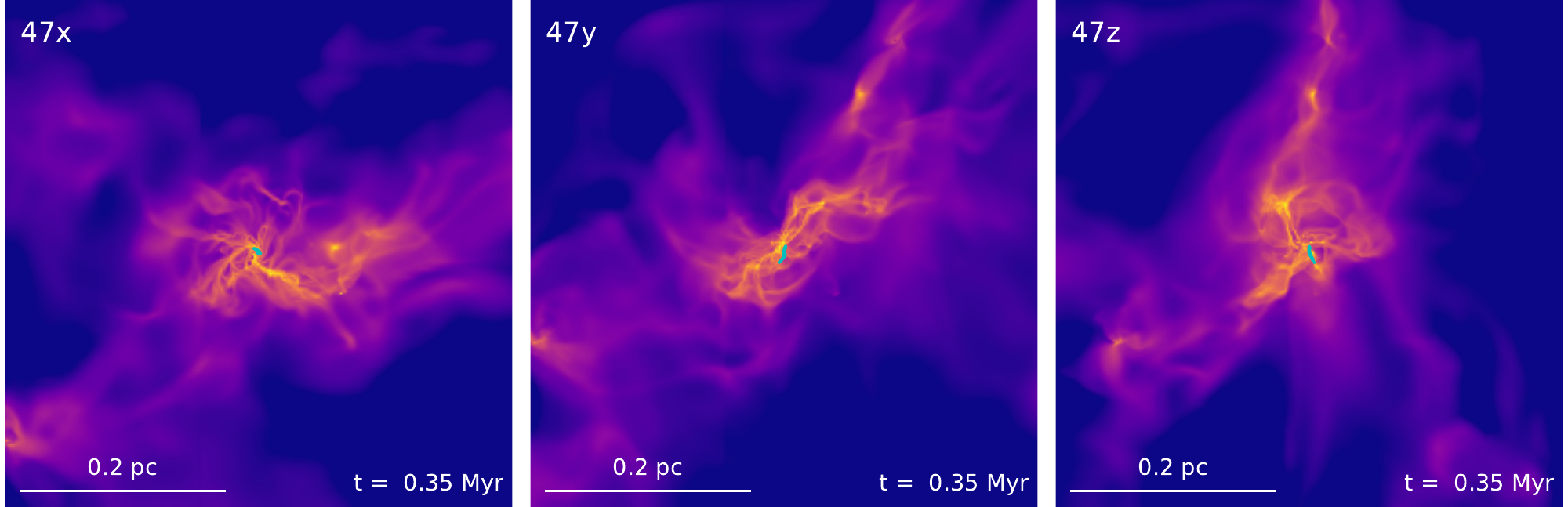}
\includegraphics[width=\hsize]{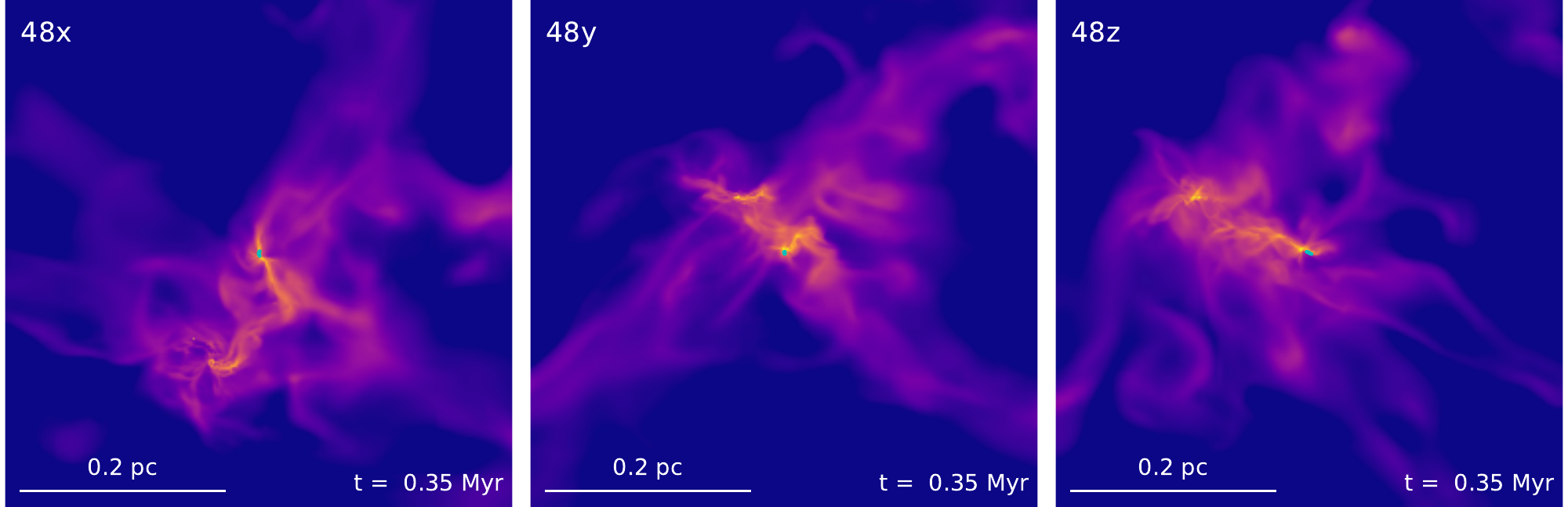}
\end{figure*}

\begin{figure*}
\centering
\includegraphics[width=\hsize]{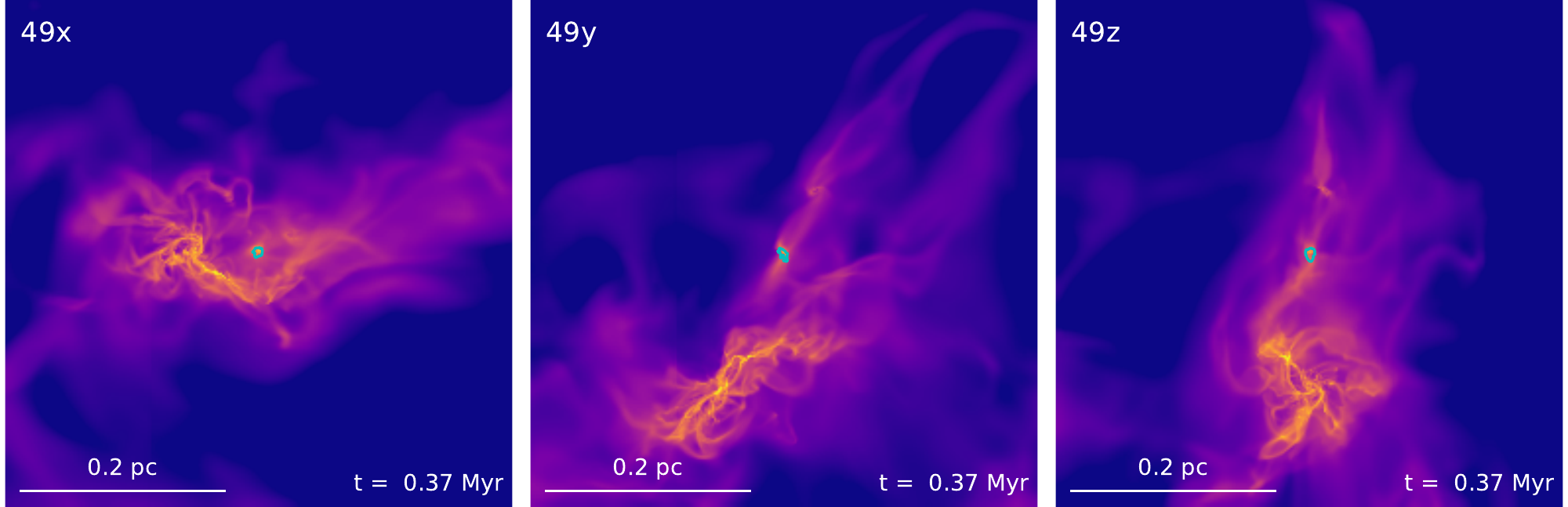}
\includegraphics[width=\hsize]{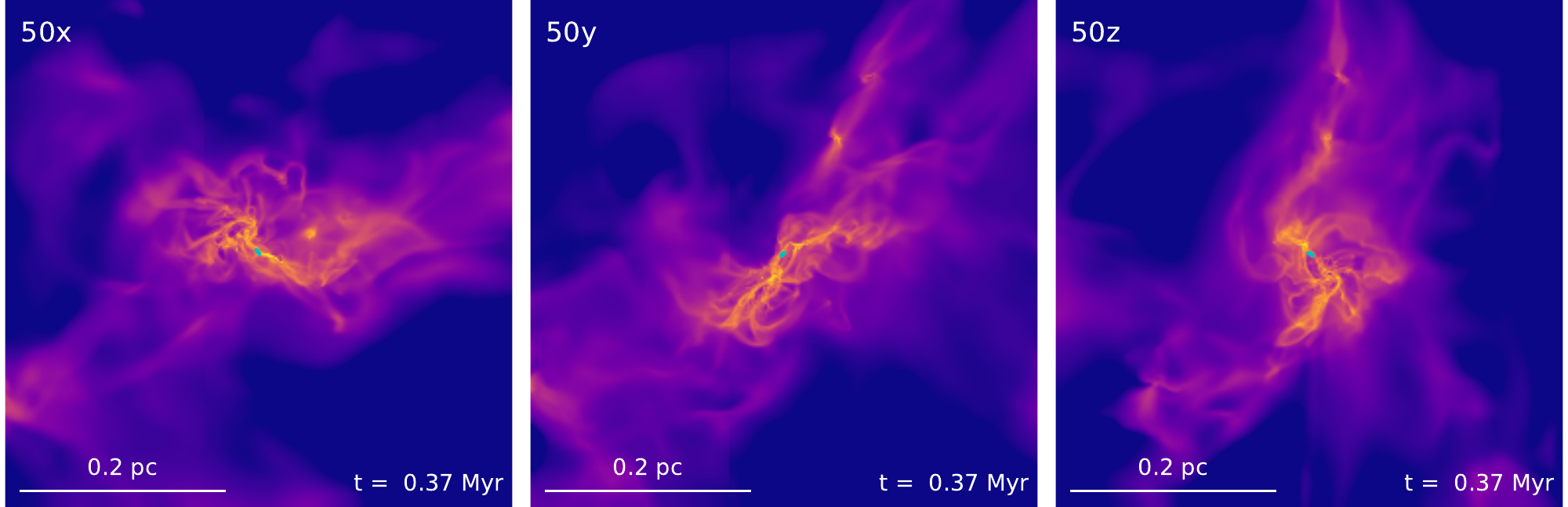}
\end{figure*}

\end{appendix}

\bsp	
\label{lastpage}
\end{document}